\documentclass[11pt,twoside]{article}
\usepackage{cal10}

\contact{Elena Pancino}
\email{elena.pancino@oabo.inaf.it}

\begin{document}

%-----------------------------------------------------------------------
%		            Paper Title 
%-----------------------------------------------------------------------
% Enter the title of the paper.
%
% EXAMPLE: \title{A Breakthrough in Astronomical Software Development}
%
% If your title is so long as to fill the page header when you print it,
% then please supply a short form as a \titlemark.  You don't need a
% \titlemark macro if your title fits in the page header.
%
% EXAMPLE:
%  \title{Rapid Development for Distributed Computing, with Implications
%         for the Virtual Observatory}
%  \titlemark{Rapid Development for Distributed Computing}

\title{An insight into the flux calibration of Gaia G-band images and BP/RP
spectrophotometry}
\titlemark{Gaia flux calibration}

%-----------------------------------------------------------------------
%		          Authors of Paper
%-----------------------------------------------------------------------
% Enter the authors followed by their affiliations.  The \author and
% \affil commands may appear multiple times as necessary.  List each
% author by giving the first name or initials first followed by the
% last name.  Authors with the same affiliations should grouped
% together. 
%
% Try to limit the front matter to no more than three \author
% commands.  Group authors with the same affiliations.  Too many
% \author commands fills the first page of the paper with little
% actual text.
%
% EXAMPLE:
% \author{S. Djorgovski\altaffilmark{1,2} and Ivan R. King}
% \affil{Astronomy Department, University of California,
%     Berkeley, CA 94720}
%
% \author{C. D. Biemesderfer\altaffilmark{3}}
% \affil{National Optical Astronomy Observatories, Tucson, AZ 85719}

\author{E. Pancino}
\affil{INAF -- Bologna Observatory, Via Ranzani 1, I-40127 Bologna, Italy}

% \author{}
% \affil{}

% Notice that some of these authors have alternate affiliations, which
% are identified by the \altaffilmark after each name. The actual alternate
% affiliation information is typeset in footnotes at the bottom of the
% first page, and the text itself is specified in \altaffiltext commands.
% There is a separate \altaffiltext for each alternate affiliation
% indicated above.
%
% EXAMPLE:
% \altaffiltext{1}{Visiting Astronomer, Cerro Tololo Inter-American Observatory. 
% CTIO is operated by AURA, Inc.\ under cooperative agreement with the National
% Science Foundation} 
% \altaffiltext{2}{Society of Fellows, Harvard University} 
% \altaffiltext{3}{Patron, Alonso's Bar and Grill}

%-----------------------------------------------------------------------
%		      Author Index Specification
%-----------------------------------------------------------------------
% Specify how each author name should appear in the author index.  The 
% \paindex{ } should be used to indicate the primary author, and the
% \aindex for all other co-authors.  You MUST use the following
% syntax: 
%
% SYNTAX:  \aindex{LASTNAME, F. M.}
% 
% where F is the first initial and M is the second initial (if
% used).  This guarantees that authors that appear in multiple papers
% will appear only once in the author index.  
%
% EXAMPLE:
% \paindex{Djorgovski, S.}
% \aindex{King, I. R.}
% \aindex{Biemesderfer, C. D.}

\paindex{Pancino, E.}
% \aindex{}
% \aindex{}

%-----------------------------------------------------------------------
%                     Author list for page header
%-----------------------------------------------------------------------
% Please supply a list of author last names for the page header. in
% one of these formats:
%
% EXAMPLES:
% \authormark{LASTNAME}
% \authormark{LASTNAME1 \& LASTNAME2}
% \authormark{LASTNAME1, LASTNAME2, ... \& LASTNAMEn}
% \authormark{LASTNAME et al.}
%
% Use the "et al." form in the case of seven or more authors, or if
% the preferred form is too long to fit in the header.

\authormark{Pancino}

% The abstract is entered in a LaTeX "environment", designated with paired
% \begin{abstract} -- \end{abstract} commands. Other environments are
% identified by the name in the curly braces.

% Poster authors ONLY may omit the abstract in order to gain a little
% more page space for the text of the poster.

\begin{abstract}

The Gaia mission is described, focussing on those technical aspects that are
necessary to understand the details of its external (absolute) flux calibration.
On board of Gaia there will be two (spectro)photometers, the blue one (BP) and the
red one (RP) covering the range 330-1050 nm, and the white light (G-band) imager
dedicated to astrometry. Given the fact that the focal plane of Gaia will be
constituted by 105 CCDs and the sources will cross the the focal plane at constant
speed, at different positions in each of the foreseen passages (on average 70--80,
but up to 350) in the mission lifetime, the ``simple" problem of calibrating the
integrated BP/RP and G-band magnitudes and the low resolution BP/RP spectra flux
turns into a very delicate and complicated issue, including CTI effects, LSF
variations across the focal plane and with time, CCD gating to avoid saturation
and the like. The calibration model requires a carefully selected set of
$\simeq$200 SpectroPhotometric Standard Stars (SPSS) with a nominal precision of a
few \%, with respect to Vega.

\end{abstract}

%-----------------------------------------------------------------------
%			Subject Index keywords
%-----------------------------------------------------------------------
% Enter up to 6 keywords describing your paper.  These will NOT be
% printed as part of your paper; however, they will be used to
% generate the subject index for the proceedings.  There is no
% standard list; however, you can consult the indices for past Calibration
% Workshop Proceedings. 


%-----------------------------------------------------------------------
%			      Main Body
%-----------------------------------------------------------------------
% Place the text for the main body of the paper here.  You should use
% the \section command to label the various sections; use of
% \subsection is optional.  Significant words in section titles should
% be capitalized.  Sections and subsections will be numbered
% automatically.

\section{The Gaia mission}

Gaia is a cornerstone mission of the ESA Space Program, presently scheduled for
launch in 2012. The Gaia satellite will perform an all-sky survey to obtain
parallaxes and proper motions to $\mu as$ precision for about 10$^9$ point-like
sources and astrophysical parameters ($T_{\mathrm{eff}}$, $\log g$, $E(B-V)$,
metallicity etc.) for stars down to a limiting magnitude of $V\simeq 20$, plus
2-30 km/s  accuracy (depending on spectral type), radial velocities for several
millions of stars down to $V < 17$. 

Such an observational effort has been compared to the mapping of the human genome
for the amount of collected data and for the impact that it will have on all
branches of astronomy and astrophysics. The expected end-of-mission astrometric
accuracies are almost 100 times better than the HIPPARCOS dataset (see Perryman et
al. 1997). This exquisite precision will allow a full and detailed reconstruction
of the 3D spatial structure and 3D velocity field of the Milky Way galaxy within
$\simeq 10$ kpc from the Sun. This will provide answers to long-standing questions
about the origin and evolution of our Galaxy, from a quantitative census of its
stellar populations, to a detailed characterization of its substructures (as, for
instance, tidal streams in the Halo, see Ibata \& Gibson, 2007, Sci. Am., 296,
40), to the distribution of dark matter. 

The accurate 3D motion of more distant Galactic satellites (as globular clusters
and the Magellanic Clouds) will be also obtained by averaging the proper motions
of many thousands of member stars: this will provide an unprecedented leverage to
constrain the mass distribution of the Galaxy and/or non-standard theories of
gravitation. Gaia will determine direct geometric distances to essentially any
kind of standard candle currently used for distance determination, setting the
whole cosmological distance scale on extremely firm bases. 

As challenging as it is, the processing and analysis of the huge data-flow
incoming from Gaia is the subject of thorough study and preparatory work by the
Data Processing and Analysis Consortium (DPAC), in charge of all aspects of the
Gaia data reduction. The consortium comprises more than 400 scientists from 25
European institutes. Gaia is usually described as a self-calibrating mission, but
it also needs {\em external} data to fix the zero-point of the magnitude system
and radial velocities, and to calibrate the classification/parametrization
algorithms. {\em All these additional data are termed auxiliary data and have to
be available, at least in part, three months before launch.} While part of the
auxiliary data already exist and must only be compiled from archives, this is not
true for several components. To this aim a coordinated programme of ground-based
observations is being organized by a dedicated inter-CU committee (GBOG), that
promotes sinergies and avoids duplications of effort.

\subsection{Science goals and capabilities}

\begin{table}[t]
\caption{Expected numbers of specific objects observed by Gaia.}
\label{tab_numbers}
\begin{center}
\begin{tabular}{l r l r}
Type & Numbers&Type & Numbers\\
\tableline
Extragalactic supernovae & 20\,000 & Extra-solar planets & 15\,000\\ 
Resolved galaxies & 10$^6$--10$^7$ & Disk white dwarfs & 200\,000\\
Quasars & 500\,000 & Astrometric microlensing events & 100 \\ 
Solar system objects & 250\,000 & Photometric microlensing events & 1000 \\ 
Brown dwarfs & $\geq$50\,000 & Resolved binaries (within 250 pc) & 10$^7$ \\ 
\end{tabular}
\end{center}
\end{table}

Gaia will measure the positions, distances, space motions, and many physical
characteristics of some billion stars in our Galaxy and beyond. For many years,
the state of the art in celestial cartography has been the Schmidt surveys of
Palomar and ESO, and their digitized counterparts.   The measurement precision,
reaching a few  millionths of a second of arc, will be unprecedented. Some
millions of stars will be measured with a distance accuracy of better than 1 per
cent; some 100 million or more to better than 10 per cent.  Gaia's resulting
scientific harvest is of almost inconceivable extent and implication. 

Gaia will provide detailed  information on stellar evolution and star formation in
our Galaxy. It will clarify the origin and formation history  of our Galaxy. The
results will precisely identify relics of tidally-disrupted accretion debris,
probe the distribution of dark matter, establish the luminosity function for
pre-main sequence stars, detect and categorize rapid evolutionary stellar phases,
place unprecedented constraints on the age, internal structure and evolution of
all stellar types, establish a rigorous distance scale framework throughout the
Galaxy and beyond, and classify star formation and kinematical and dynamical
behaviour within the Local Group of galaxies. 

Gaia will pinpoint exotic objects in colossal and almost unimaginable numbers:
many thousands of extra-solar  planets will be discovered (from both their
astrometric wobble and from photometric transits) and their detailed orbits and
masses determined; tens of thousands of brown dwarfs and white dwarfs will be
identified; tens of  thousands of extragalactic supernovae will be discovered;
Solar System studies will receive a massive impetus through the observation of
hundreds of thousands of minor planets; near-Earth objects, inner Trojans and
even  new trans-Neptunian objects, including Plutinos, may be discovered. 

Gaia will follow the bending of star light by the Sun and major planets over the
entire celestial sphere, and  therefore directly observe the structure of
space-time -- the accuracy of its measurement of General Relativistic light
bending may reveal the long-sought scalar correction to its tensor form. The PPN
parameters $\gamma$ and $\beta$, and the solar quadrupole moment J2, will be
determined with unprecedented precision. All this, and more, through the accurate
measurement of star positions. 

We summarize some of the most interesting object classes that will be observed by
Gaia, with estimates of the expected total number of objects, in
Table~\ref{tab_numbers}. For more information on the Gaia mission:
http://www.rssd.esa.int/Gaia. More information for the public on Gaia and its
science capabilities are contained in the {\em Gaia information
sheets}\footnote{http://www.rssd.esa.int/index.php?pro
ject=GAIA\&page=Info\_sheets\_overview.}. An excellent review of the science
possibilities opened by Gaia can be found in Perryman et al. (1997).

\subsection{Launch, timeline and data releases}

\begin{figure}
\epsscale{0.75}
\plotone{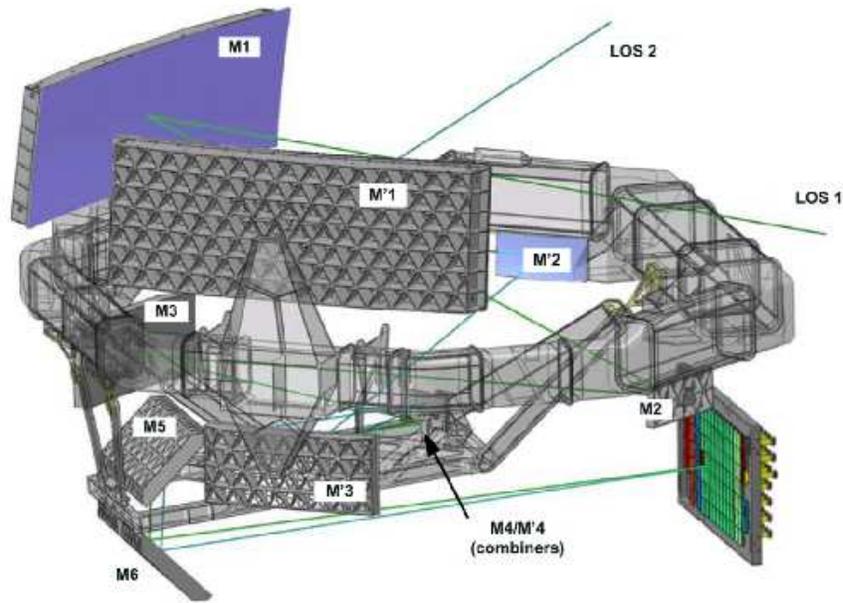}
\caption{The two Gaia telescopes, mounted on a compact torus, point towards two
lines of sight separated by 106.5$^{\rm o}$, and converging on the same focal
plane. \copyright ESA}
\label{pancino_fig_tels}
\end{figure}

The first idea for Gaia began circulating in the early 1990, culminating in a
proposal for a  cornerstone mission within ESA's science programme submitted in
1993, and a workshop in Cambridge in June 1995. By the time the final catalogue
will be released approximately in 2020, almost two decades of work will have
elapsed between the orginal concept and mission completion.  

Gaia will be launched by a Soyuz carrier (rather than the initially foreseen
Ariane 5) in 2012 from French Guyana and will start operating once it will reach
its Lissajus orbit around L2 (the unstable Langrange point of the Sun and
Eart-Moon system), in about one month. Two ground stations will receive the
compressed Gaia data during the 5 years\footnote{If -- after careful evaluation --
the scientific output of the mission will benefit from an extension of the
operation period, the satellite should be able to gather data for one more year,
remaining within the Earth eclipse.} of operation: Cebreros (Spain) and Perth
(Australia). The data will then be transmitted to the main data centers throughout
Europe to allow for data processing. We are presently in technical development
phase C/D, and the hardware is being built, tested and assembled. Software
development started in 2006 and is presently producing and testing pipelines with
the aim of delivering to the astrophysical community a full catalogue and dataset
ready for scientific investigation.

Apart from the end-of-mission data release, foreseen around 2020, some
intermediate data releases are foreseen. In particular, there should be one first
intermediate release covering either the first 6 months or the first year of
operation, followed by a second and possibly a third intermediate release, that
are presently being discussed. The data analysis will proceed in parallel with
observations, the major pipelines re-processing all the data every 6 months, with
secondary cycle pipelines -- dedicated to specific tasks -- operating on different
timescales. In particular, verified science alerts, based on unexpected
variability in flux and/or radial velocity, are expected to be released within 24
hours from detection, after an initial period of testing and fine-tuning of the
detection algorithms. 

\subsection{Mission concepts}

\begin{figure}
\plottwo{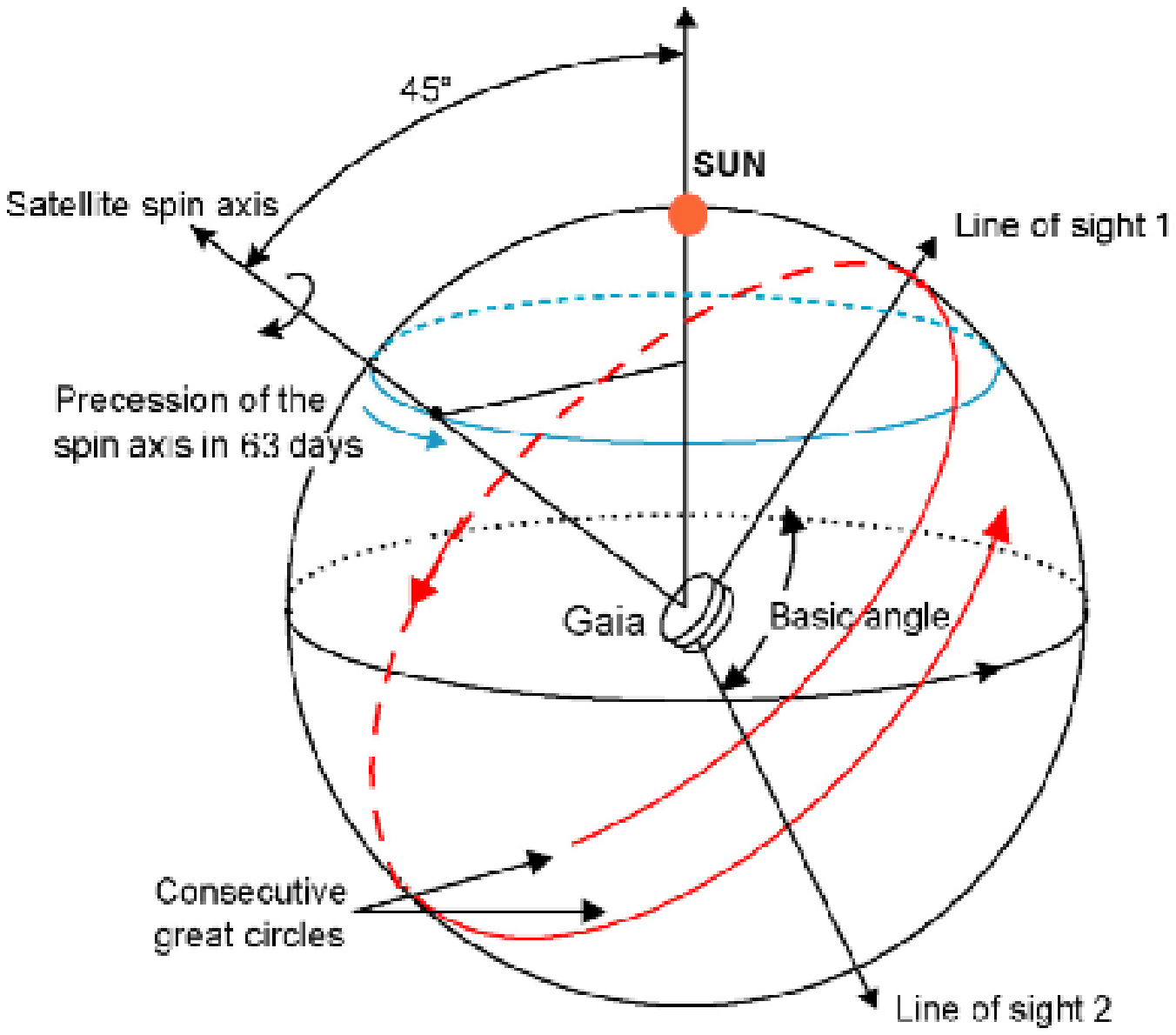}{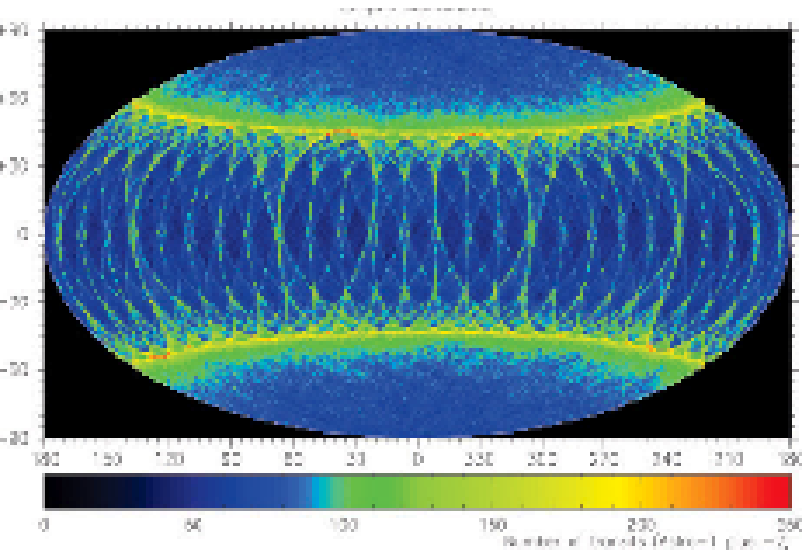}
\caption{Left: the scanning law of Gaia during main operations; Right: the average
number of passages on the sky, in ecliptic coordinates. \copyright ESA}
\label{pancino_fig_scan}
\end{figure}

During its 5-year operational lifetime, the satellite will continuously spin
around its axis, with a constant speed of 60~arcsec/sec. As a result, over a
period of 6 hours, the two astrometric fields of view will scan across all objects
located along the great circle perpendicular to the spin axis
(Figure~\ref{pancino_fig_scan}, left panel). As a result of the basic angle of
106.5$^{\rm o}$ separating the astrometric fields of view on the sky
(Figure~\ref{pancino_fig_tels}), objects transit the second field of view with a
delay of 106.5 minutes compared to the first field. Gaia's spin axis does not
point to a fixed direction in space, but is carefully controlled so as to precess
slowly on the sky. As a result, the great circle that is mapped by the two fields
of view every 6 hours changes slowly with time, allowing reapeated full sky
coverage over the mission lifetime. The best strategy, dictated by thermal
stability and power requirements, is to let the spin axis precess (with a period
of 63 days) around the solar direction with a fixed angle of 45$^{\rm o}$. The
above scanning strategy, referred to as ``revolving scanning", was successfully
adopted during the Hipparcos mission. 

Every sky region will be scanned on average 70-80 times, with regions lying at
$\pm$45$^{\rm o}$ from the Ecliptic Poles being scanned on average more often than
other locations. Each of the Gaia targets will be therefore scanned (within
differently inclined great circles) from a minimum of approximately 10 times to a
maximum of 250 times (Figure~\ref{pancino_fig_scan}, right panel). Only point-like
sources will be observed, and in some regions of the sky, like the Baade's window,
$\omega$ Centauri or other globular clusters, the star density of the two combined
fields of view will be of the order of 750\,000 or more per square degree,
exceeding the storage capability of the onboard processors, so Gaia will not study
in detail these dense areas. 

\begin{figure}
\plotone{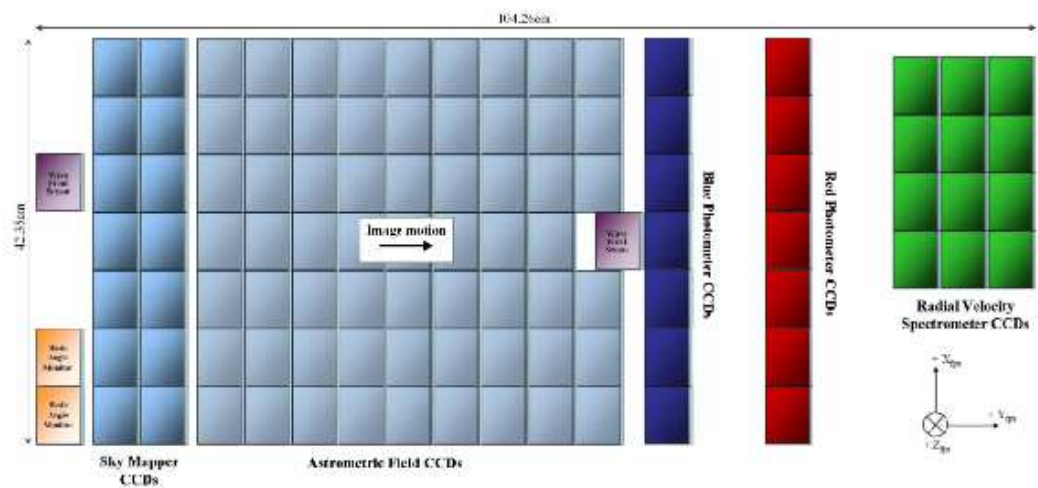}
\caption{The 105 on the Gaia focal plane. \copyright ESA}
\label{pancino_fig_foc}
\end{figure}

\subsection{Focal plane}

Figure~\ref{pancino_fig_foc} shows the focal plane of Gaia, with its 105 CCDs,
which are read in TDI (Time Delay Integration) mode: objects enter the focal plane
from the left and cross one CCD in 4 seconds. Apart from some technical CCDs that
are of little interest in this context, the first two CCD columns, the Sky Mappers
(SM), perform the on-board detection of point-like sources, each of the two
columns being able to see only one of the two lines of sight. After the objects
are identified and selected, small windows are assigned, which follow them in the
astrometric field (AF) CCDs where white light (or G-band) images are obtained
(Section~\ref{sec-af}). Immediately following the AF, two additional columns of
CCDs gather light from two slitless prism spectrographs, the blue
spectrophotometer (BP) and the red one (RP), which produce dispersed images
(Section~\ref{sec-phot}). Finally, objects transit on the Radial Velocity
Spectrometer (RVS) CCDs to produce higher resolution spectra around the Calcium
Triplet (CaT) region (Section~\ref{sec-rvs}). 

\subsection{Astrometry} 
\label{sec-af}

\begin{figure}
\plottwo{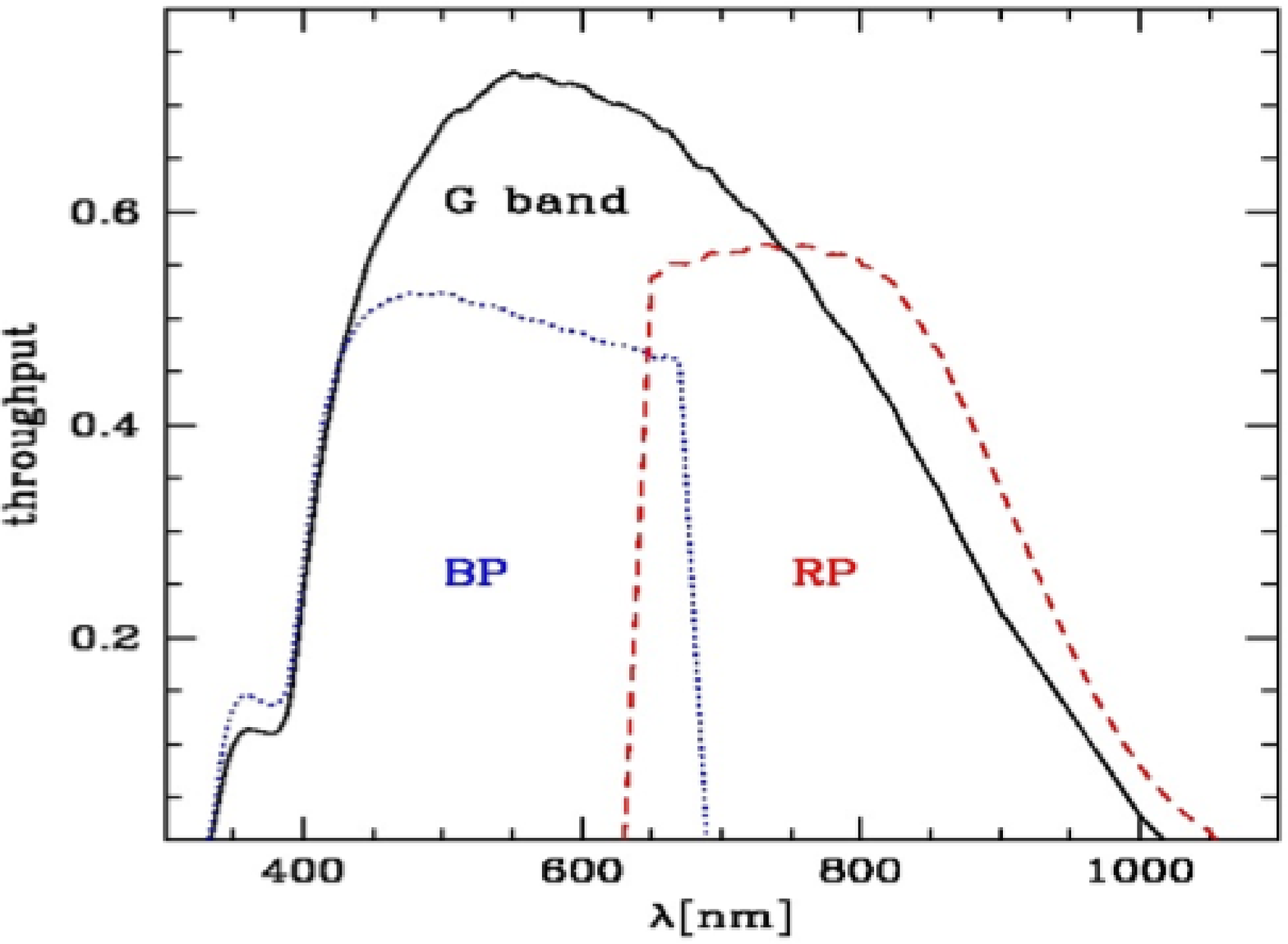}{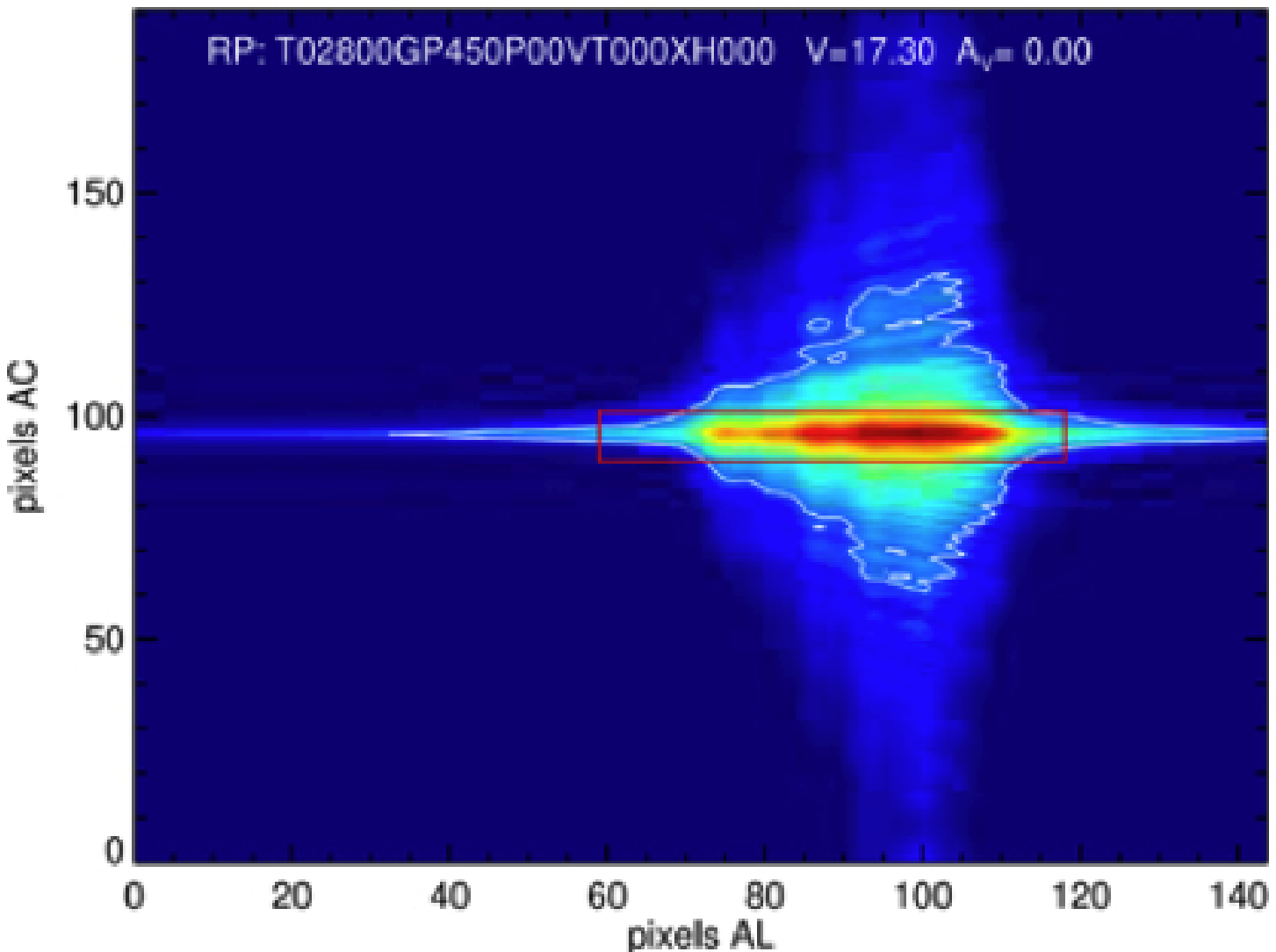}
\caption{Left: the passbands of the G-band, BP and RP; Right: a simulated RP
dispersed image, with a red rectangle marking the window assigned for compression
and ground telemetry. \copyright ESA} \label{pancino_fig_phot}
\end{figure}

The AF CCDs will provide G-band images, i.e., white light images where the
passband is defined by the telescope optics transmission and the CCDs sensitivity,
with a very broad combined passband ranging from 330 to 1050~nm and peaking around
500--600~nm (Figure~\ref{pancino_fig_phot}). The objective of Gaia's astrometric
data reduction system is the construction of core mission products: the five
standard astrometric parameters, position ($\alpha$, $\delta$), parallax
($\varpi$), and proper motion ($\mu_{\alpha}$, $\mu_{\delta}$) for all observed
stellar objects. The expected end-of-mission precision in the proper motions is
expected to be better than 10~$\mu$as for G$<$10 stars, 25~$\mu$as for G=15, and
300~$\mu$as for G=20. For parallaxes, considering a G=12 star, we can expect to
have distances at better than 0.1\% within 250~pc, 1\% within 2700~pc, and 10\%
within 10~kpc.

To reach these end-of-mission precisions, the average 70--80 observations per
target gathered during the 5-year mission duration will have to be combined into a
single, global, and self-consistent manner. 40~Gb of telemetry data will first
pass through the Initial Data Treatment (IDT) which determines the image
parameters and centroids, and then performes an object cross-matching. The output
forms the so-called One Day Astrometric Solution (ODAS), together with the
satellite attitude and calibration, to the sub-milliarcsecond accuracy.  The data
are then written to the Main Database. 

The next step is the Astrometric Global Iterative Solution (AGIS) processing. AGIS
processes together the attitude and calibration parameters with  the source
parameters, refining them in an iterative procedure that stops when the
adjustments become sufficiently small. As soon as new data come in, on the basis
of 6 months cycles, all the data in hand are reprocessed toghether from scratch.
This is the only scheme that allows for the quoted precisions, and it is also the
philosophy that justifies Gaia as a self-calibrationg mission. The primary AGIS
cycle will treat only stars that are flagged as single and non-variable (expected
to be around 500 millions), while other kinds of objects will be computed in
secondary AGIS cycles that utilize the main AGIS sulution. Dedicated pipelines for
specific kinds of objects (asteroids, slightly extended objects, variable objects
and so on) are being put in place to extract the best possible precision. Owing to
the large data volume (100~Tb) that Gaia will produce, and to the iterative nature
of the processing, the computing challenges are formidable: AGIS processing alone
requires some 10$^{21}$~FLOPs which translates to runtimes of months on the ESAC
computers in Madrid.

\subsection{Spectrophotomety}
\label{sec-phot}

The primary aim of the photometric instrument is mission critical in two respects:
(i) to correct the measured centroids position in the AF for systematic chromatic
effects, and (ii) to classify and determine astrophysical characteristics of all
objects, such as temperature, gravity, mass, age and chemical composition (in the
case of stars).

The BP and RP spectrophotometers are based on a dispersive-prism approach such
that the incoming light is not focussed in a PSF-like spot, but dispersed along
the scan direction in a low-resolution spectrum. The BP operates between
330--680~nm while the RP between 640-1000~nm (Figure~\ref{pancino_fig_phot}). Both
prisms have appropriate broad-band filters to block unwanted light. The two
dedicated CCD stripes cover the full height of the AF and, therefore, all objects
that are imaged in the AF are also imaged in the BP and RP. 

The resolution is a function of wavelength, ranging from 4 to 32 nm/pix for BP and
7 to 15 nm/pix for RP. The spectral resolution, R=$\lambda/\delta \lambda$ ranges
from 20 to 100 approximately. The dispersers have been designed in such a way that
BP and RP spectra are of similar sizes (45 pixels). Window extensions meant to
measure the sky background are implemented. To compress the amount of data
transmitted to the ground, all the BP and RP spectra -- except for the brightest
stars -- are binned on chip in the across-scan direction, and are transmitted to
the ground as one-dimensional spectra. Figure~\ref{pancino_fig_phot} shows a
simulated RP spectrum, unbinned, before windowing, compression, and telemetry.

The final data products will be the end-of-mission (or intermediate releases) of
global, combined BP and RP spectra and integrated magnitudes M$_{BP}$ and
M$_{RP}$. Epoch spectra will be released only for specific classes of objects,
such as variable stars and quasars, for example. The internal flux calibration of
integrated magnitudes, including the M$_G$ magnitudes as well, is expected at a
precision of 0.003~mag for G=13 stars, and for G=20 stars goes down to 0.07~mag in
M$_G$, 0.3~mag in M$_{BP}$ and M$_{RP}$. The external calibration should be
performed with a precision of the order of a few percent (with respect to Vega).

\begin{figure}
\plotone{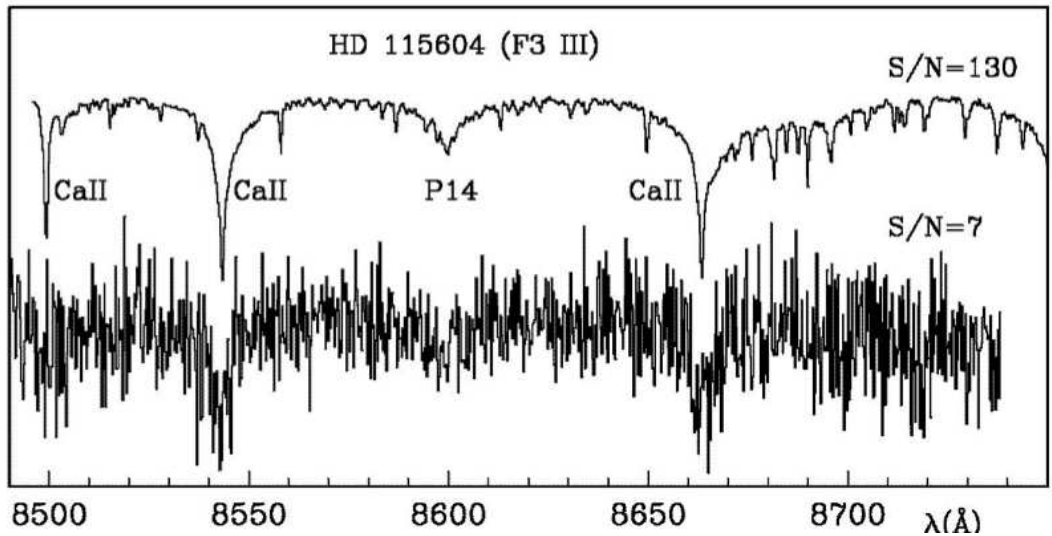}
\caption{Simulated RVS end-of-mission spectra for the extreme cases of 1 single 
transit (bottom spectrum) and of 350 transits (top spectrum). \copyright ESA}
\label{pancino_fig_rvs}
\end{figure}

\subsection{High-resolution spectroscopy}
\label{sec-rvs}

The primary objective of the RVS is the acquisition of radial velocities, which
combined with positions, proper motions, and parallaxes will provide the means to
decipher the kinematical state and dynamical history of our Galaxy.

The RVS will provide the radial velocities of about 100--150 million stars up to
17-th magnitude with precisions ranging  from 15 km s$^{-1}$ at the faint end, to
1 km s $^{-1}$ or better at the bright end. The spectral resolution,
R=$\lambda$/$\delta\lambda$ will be 11\,500. Radial velocities will be obtained by
cross-correlatinfg observed spectra with either a template or a mask. An initial
estimate of the source atmospheric parameters will be used to  select  the  most
appropriate template or mask. On average, 40 transits will be collected for each
object during the 5-year lifetime of Gaia, since the RVS does not cover the whole
width of the Gaia AF (Figure~\ref{pancino_fig_foc}). In total, we expect to obtain
some 5 billion spectra (single transit) for the brightest stars. The analysis of
this huge dataset will be complicated, not only because of the sheer data volume,
but also because the spectroscopic data analysis relies on the multi-epoch
astrometric and photometric data. 

The covered wavelength range (847-874 nm) (Figure~\ref{pancino_fig_rvs}) is a rich
domain, centered on the infrared calcium triplet: it  will not only provide radial
velocities, but also many stellar and interstellar diagnostics. It has been
selected to coincide with the energy distribution peaks of G anf K type stars,
which are the most abundant targets. In early type stars, RVS spectra may contain
also weak Helium lines and N, although they will be dominated by the Paschen
lines. The RVS data will effectively complement the astrometric and photometric
observations, improving object classification. For stellar objects, it will
provide atmospheric parameters such as effective temperature, surface gravity, and
individual abundances of key elements such as Fe, Ca, Mg, Si for millions of stars
down to G$\simeq$12. Also, Diffuse Intertellar Bands (DIB) around 862 nm will
enable the derivation of a 3D map of interstellar reddening. 

\subsection{The DPAC}
\label{sec-dpac}

ESA will take care of the satellite design, build and testing phases, of launch
and operation, and of the data telemetry to the ground, managing the ESAC
datacenter in Madrid, Spain. The data treatment and analysis is instead
responsibility of the European scientific community. In 2006, the announcement of
opportunity opened by ESA was successfully answered by the Data Processing and
Analysis Consortium (DPAC), a consortium that is presently counting more than 400
scientists in Europe (and outside) and more than 25 scientific institutions. 

The DPAC governing body, or executive (DPACE) oversees the  DPAC activities and 
the work has been organized among a few Coordination Units (CU) in charge of
different aspects of data treatment:

\begin{itemize}
\item{{\bf CU1. System Architecture} (manager:  O' Mullane), dealing with all
aspects of hardware and software, and coordinating  the framework for software
develepment and data management.}
\item{{\bf CU2. Data Simulations} (manager: Luri), in charge of the simulators of
various stages of data products, necessary for software development and testing.}
\item{{\bf CU3. Core processing} (manager: Bastian), developing the main pipelines
such as IDT, AGIS and astrometry processing in general.}
\item{{\bf CU4. Object Processing} (managers: Pourbaix/Tanga), for the processing
of objects that require special treatment such as minor bodies of the Solar
system, for example.}
\item{{\bf CU5. Photometric processing} (manager: van Leeuwen), dedicated to the
BP, RP, and M$_G$ processing and calibration, including image reconstruction,
background treatment, and crowding treatment, among others.}
\item{{\bf CU6. Spectroscopic Processing} (managers: Katz/Cropper), dedicated to
RVS processing and radial velocity determination.}
\item{{\bf CU7. Variability Processing} (managers: Eyer/Evans/Dubath), dedicated
to processing, classification and parametrization of variable objects.}
\item{{\bf CU8. Astrophysical Parameters} (managers: Bailer-Jones/Thevenin),
developing object classification software and, for each object class, software for
the determination of astrophysical parameters.}
\item{{\bf CU9. Catalogue Production and Access} (to be activated in the near
future), responsible for the production of astrophysical catalogues and for the 
publication of Gaia data to the scientific community. }
\end{itemize}

These are flanked by a few working groups (WG) that deal with aspects that are
either transversal among the various CUs (such as the GBOG, coordinating the
ground based observations for the external calibration of Gaia) or  of general
interest (such as the Radiation task force, serving as the interface between DPAC
and the industry in all matters related to CCD radiation tests). 

\section{The flux calibration of Gaia data}

Calibrating (spectro)photometry obtained from the usual type of ground based
observations (broadband imaging, spectroscopy) is not a trivial task, but  the
procedures are well known (see e.g., Bessell, 1999) and several scientists have
developed sets of standard stars appropriated for the more than 200 photometric
systems known, and for spectroscopic observations. Generally, magnitudes are
calibrated to a standard system with equations in the form

$$ M = m + ZP + \alpha(colour) + \beta(airmass) $$

where M is the calibrated magnitude in a chosen photometric band, m the observed
(instrumental) one in the same (or very similar) band, $\alpha$ is the colour term
and $\beta$ the extinction coefficient, due the Earth atmospheric extinction. For
the spectra, usually the instrumental effect on the observed sepctral energy
distribution (SED) is parametrized as

$$ S_{obs} (\lambda) = R(\lambda)~ S(\lambda) $$

where the observed SED, S$_{obs} (\lambda)$, is the result of the convolution of
the ``true'' SED, S($\lambda$), with all the instrumental (transmissivity, quantum
efficiencies) effects, which are empyrically determined in the form of a response
curve R($\lambda$) through the use of spectrophotometric standard stars (SPSS). In
the case of Gaia, several instrumental effects -- much more complex than those
usually encountered -- redistribute light along the SED of the observed objects.
In particular: the TDI integration mode, the fact that the focal plane is so
large, the radiation damage and resulting CTI (charge transfer inefficiencies),
the  fact that the whole instrumental model is well known only before launch. 

\subsection{Challenges}

The most difficult Gaia data to calibrate are the BP and RP spectra, requiring a
new approach to the derivation of the calibration model
(Section~\ref{pancino_sec_dm}) and to the SPSS needed to perform the actual
calibration (Section~\ref{pancino_sec_spss}). The large focal plane with its large
number of CCDs makes it so that different observations of the same star will be
generally on different CCDs, with different quantum efficiencies. Also, each CCD
is in a different position, with different optical distorsions, optics
transmissivity and so on. Therefore, each wavelength and each position across the
focal plane has its  (sometimes very different) PSF (point spread function). The
TDI and continuous reading mode, combined with the need of compressing the data
before on-ground transmission, make it necessary to translate the full PSF into a
linear (compressed into 1D) LSF (line spread function), which of course add
complication into the picture. In-flight instrument monitoring is foreseen, but
never comparable to the full characterization that will be performed before
launch, so the real instrument -- at a  certain observation time -- will be
different from the theoretical one assumed initially, and this difference will
change with time.

Special mention deserves the radiation damage, one of the most important factors
in the time variation of the instrument model. It has particular impact onto the
BP and RP dispersed images since the objects travel along the BP and RP CCD strips
in a direction that is parallel to the spectral dispersion (wavelength
coordinate). Radiation damage causes traps that subtract photons from each passing
object at a position corresponding to a certain wavelength. Slow traps release the
trapped charges once the object is already passed, while fast traps can realease
the  charges within the same object, but at a different wavelenght. Given the low
resolution, one pixel can cover as much as 15--20~nm (depending on the wavelength)
and therefore the net effect of radiation damage can be to alter significantly the
SED of some spectra. Possible solutions under testing are the equivalent of CCD
pre-flashing, the statistical modeling of the traps behaviour and the fact that
different transits for the same object will be affected differently by CTI
effects, allowing for a certain degree of correction through average or median
spectra. Finally the PSF/LSF itself is generally larger than one Gaia pixel in the
BP/RP spectra, introducing a large LSF smearing effect, i.e., the spread of
photons with one particular wavelength into a large range of wavelengths.  

In this paper, we will adopt the current Gaia calibration philosophy, where most
of these intrumental effects are taken into account during the so-called
internal flux calibration. A large number of well behaved stars ({\em internal
standards}) observed by Gaia will be used to report all observations to a {\em
reference} instrument, on the same instrumental flux and wavelength scales. All
transits for each object observed by Gaia will be then averaged to produce one
single BP and RP spectrum for each object, with its  integrated instrumental
magnitudes: M$_G$, M$_{BP}$, and M$_{RP}$. Only for specific classes of objects,
epoch spectra and magnitudes will be released, with variable stars as an obvious
example. The mean and epoch spectra will be mostly free from many of the
problems examined just above, but they will still contain residuals due to the 
imperfect knowledge of the real instrument at each precise moment of time, and
the most significant effects are expected to be the LSF smearing and the CTI
effects.

In this paradigm, the internal and external flux and wavelength calibrations are
treated as two entirely  separated and consecutive pieces of the CU5 photometric
pipeline, with different calibration models. So, from next Section, we start
always from internally calibrated BP/RP spectra and M$_G$, M$_{BP}$, and
M$_{RP}$ magnitudes, without giving importance to the exact way they are
produced. Presently, two alternative approaches are being considered to maximize
the precision of the global calibration procedure: the first one is a {\em
hybrid model} that partially combines internal and external models (Montegriffo
et al. 2010), while the second is the so-called {\em full forwarding model}
(Carrasco et al. 2010, in preparation), using the same calibration model for
both the internal and external calibration.

\subsection{The external calibration teams}

Two Development units (DU) within CU5 (Photometric processing), are dedicated to
the external calibration of Gaia photometry.  They are DU13: ``Instrument absolute
response characterisation:  ground-based preparation'' coordinated by E.~Pancino
and DU14" ``Instrument absolute response characterisation: definition and
application'' coordinated by C.~Cacciari. They are both based in Bologna, Italy,
in collaboration with the Bologna, Barcelona, and Groningen Universities. The
actual team members at the time of writing are: G. Altavilla, M. Bellazzini, A.
Bragaglia, C. Cacciari, J. M. Carrasco, G. Cocozza, L. Federici, F. Figueras, F.
Fusi Pecci, C. Jordi, S. Marinoni, P. Montegriffo, E. Pancino, S. Ragaini, E.
Rossetti, S. Trager.

\subsection{Dispersion Matrix basic definition}
\label{pancino_sec_dm}

If we concentrate now on the mean, internally calibrated BP/RP spectra
calibration, we can write: 

$$ S_{obs} (\kappa_I) = 
\sum_{\lambda_i = 0}^{N} T(\lambda_i)~L_{\lambda_i} (\kappa_I - \kappa_P
(\lambda_i))~S_{true}(\lambda_i)$$

where S$_{obs}$ and S$_{true}$ are the observed and ``true'' SEDs respectively,
expressed the first in Gaia pixels $\kappa_I$ and the second in wavelength
intervals $\lambda_i$ corresponding to the actual sampling of the SPSS used in the
flux calibration process. T($\lambda_i)$ is a combination of all the instrument
and telescope transmissivity functions and aperture, while L is the LSF at a
certain $\lambda_i$, centered at the appropriate $\kappa_I$ pixel, but of course
calculated over the whole wavelenght interval from $\lambda_i$=0 to N (the total
number of samples in the tabulated SPSS spectrum).

\begin{figure}
\plottwo{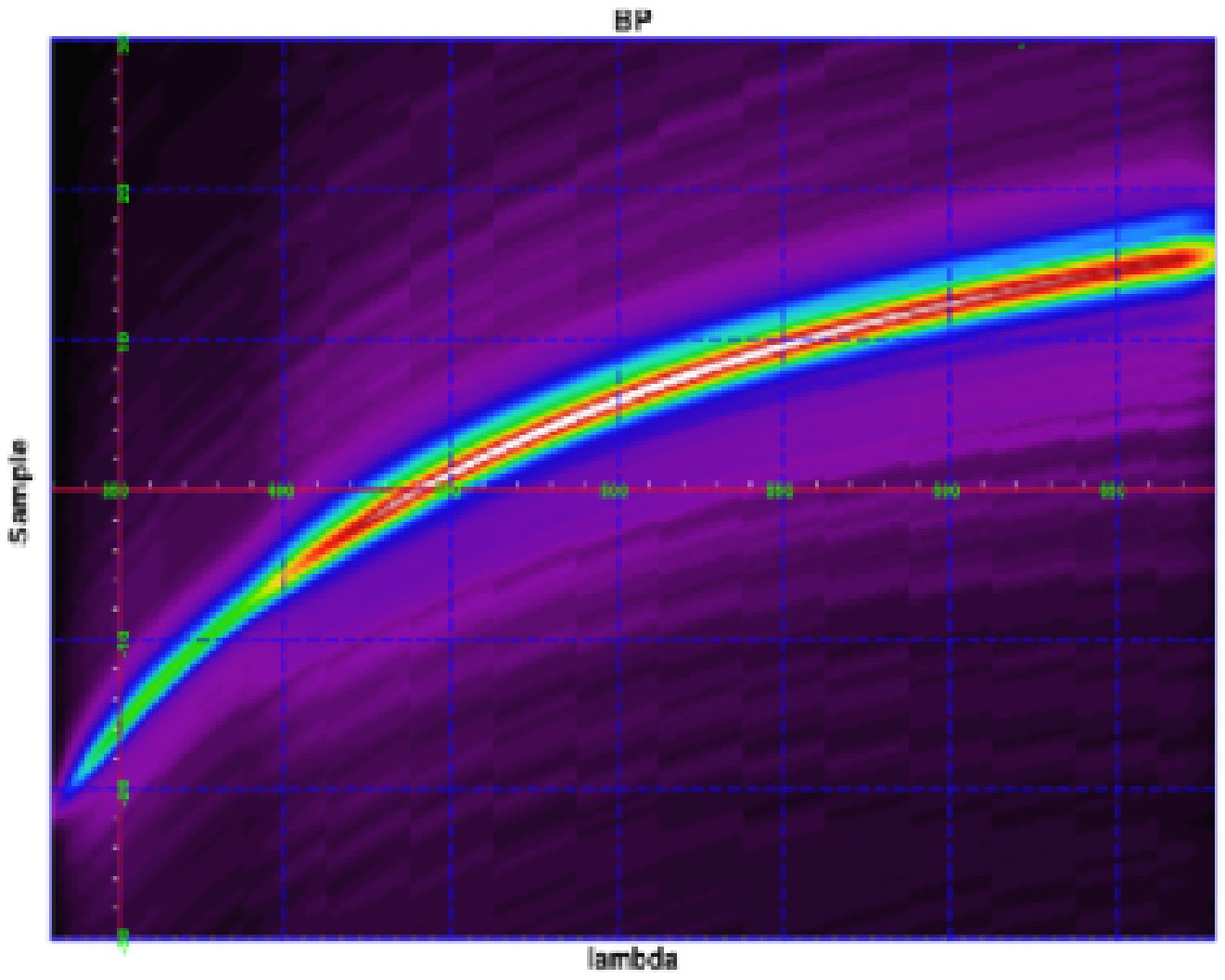}{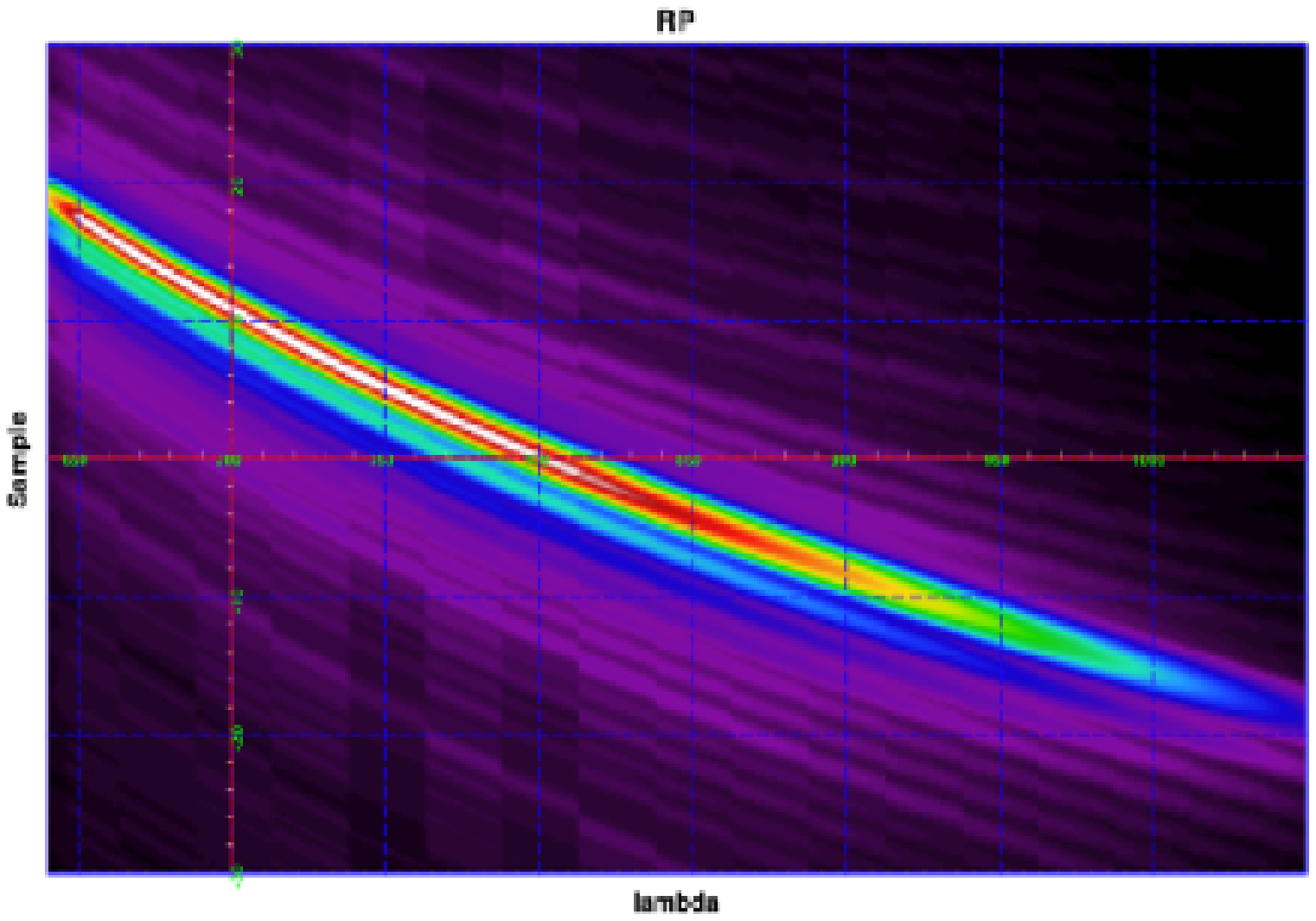}
\caption{Graphical example of a dispersion matrix D, derived with a simulated SPSS
set for the BP (left) and RP (right) instruments, on an arbitrary color scale.  \copyright ESA}
\label{pancino_fig_dm}
\end{figure}

Such an equation can be written in its much simpler matricial form:

$$ S_{obs} = D \times S_{true} $$

where D is called a ``Dispersion Matrix'', an object that can be determined if
S$_{obs}$ and S$_{true}$ are known, i.e.,  using a well defined set of SPSS
observed by Gaia, that also have well known SED (see below). Once D is properly
determined, it can be inverted to convert each mean, internally calibrated BP/RP
spectrum\footnote{Incidentally, for some object classes that need it, such as
variable stars, sigle transits -- the so called {\em epoch spectra} -- will be
published. The described calibration  model can be applied also to single epoch
spectra once they have been internally  calibrated, i.e., reported to a common,
instrumental scale of flux and wavelength.} (S$_{obs}$) into an flux calibrated
spectrum S$_{true}$

$$ S_{true} = D^{-1} \times S_{obs} $$

The main advantage of this approach is that D contains (and therefore corrects
empyrically for) the actual effects of LSF smearing -- even if the real shape of
the LSF is not perfectly known a priori. More than that, the effective LSF -- as
determined with the chosen SPSS set -- at each wavelenght can be extracted from
each column of the matrix. The matrix rows represent instead the effective
passbands corresponding to each Gaia pixel, including the full effect of LSF
smearing. This peculiar property of the dispersion matrix makes it the best (and
possibly only) solution to the external calibration of Gaia BP/RP spectra. By
definition, the dispersion matrix D contains also the actual dispersion function,
which can be seen in Figure~\ref{pancino_fig_dm} as the curved structure close to
the diagonal of the matrix. 

Finally, an important by-product of the described calibration model is the
absolute wavelength calibration of the BP/RP spectra to a precision of {\em at
least} a few tenths of a Gaia pixel\footnote{This first estimate of the wavelength
zero point and scale precision is based on a slightly outdated calibration model
formulation, an therefore has to be considered just as an upper limit to a more
realistic uncertainty (Montegriffo, 2010, private communication.} (Montegriffo \&
Bellazzini 2009b), which is automatically performed toghether with the absolute
flux calibration. 

There are a few problems in the use of the dispersion matrix as proposed. We will
discuss in the three following Sections the two most important ones and their
proposed solutions: (1) the matrix is rectangular and its inversion is not so
straightforward; (2) the matrix needs a set of independent vectors to be
determined in a non-degenerate way (which also implies that the set of SPSS must
be carefully chosen).

\subsection{Smoothing the input SPSS spectra}

Clearly, the dispersion matrix is a rectangular matrix: the Gaia observations have
a smaller number of samples (pixels) than the SPSS spectra used to build their
calibration model (wavelength sampling). Inverting a rectangular matrix is a
non-trivial task so if we want to use the inverted matrix to calibrate our data we
must find a method to reduce the dimensionality of the input SPSS flux tables. A
few different methods have been considered such as the B splines representation
(Montegriffo \& Bellazzini, 2009a), Gaussian smoothing with variable width
Gaussians (Montegriffo, 2010, private communication), and smoothing through
rectangular functions corresponding to the Gaia pixels (Montegriffo et al. 2010).
The idea is that the S$_{true}$ of each SPSS is observed from the ground with a
resolution that exceeds the Gaia BP/RP one by at least a factor of 4-5, and then
compressed in a way that minimizes information losses. Once the SPSS spectra are
smoothed, the dispersion matrix becomes a square {\em effective disperision
matrix}, D$_e$

$$ S_{obs} = D_e \times S_{smooth} $$

However, D$_e$ can still be non-diagonal or degenerate and before proceeding we
must find the criteria to build the best possible D$_e$ with the data in hand. 

\subsection{The ideal SPSS set}

One reason why the dispersion matrix is non-diagonal, is that the  SPSS adopted
set can {\em never} be an orthogonal set of independent calibrators: stars are all
similar to each other, they have all a black-body like continuum with some
features (absorption and emission lines or bands). As a result, if a dispersion
matrix is built  with a particular set of SPSS, such as white dwarfs and hot
subdwarfs (the ideal calibrators in the classic spectroscopic observations), it
will be able to calibrate properly only objects with similar spectra, i.e.,
relatively smooth, with some absorption lines in the blue part of the spectrum. 

\begin{figure}
\epsscale{0.75}
\vspace{-0.3cm}
\plotone{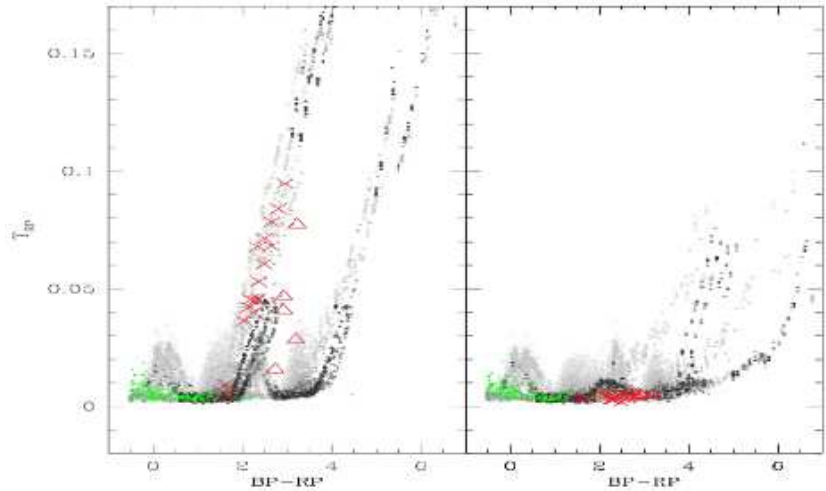}
\vspace{-0.3cm}
\caption{On both panels, grey dots are simulated spectra of different metallicity
and reddening (this explains the parallel sequences). Green  dots are white dwarfs
and hot subdwwarfs, while red symbols are different types of red stars with
significant absorption features, i.e., molecular bands. Abscissae represent the
BP--RP color, ordinates are the difference between the ``known'' magnitude of the
used SEDs and the ``calibrated'' ones obtained with a dispersion matrix. Left: all
points are calibrated with a matrix built only with hot spectral types (green
dots): the reddest stars are calibrated with an {\em error} of 0.15~mag and more.
Right: all points are calibrated with a matrix determined using also 10 red SPSS
with absorption features (red symbols): all stars with SEDs similar to the 10 red
stars (2~mag$<$BP--RP$<$4~mag) are calibrated with an {\em error} of less than
1\%. \copyright ESA}
\label{pancino_fig_red}
\end{figure}

An example of the above case is shown in Figure~\ref{pancino_fig_red}, where two
different dispersion matrices are used to calibrate the same set of smulated  Gaia
observations. In the first case (left panel of Figure~\ref{pancino_fig_red}), a
dispersion matrix is built using only white dwarfs and hot subdwarfs, with a
minority of solar type stars, and it can clearly be seen that red stars with deep
absorption bands are calibrated with an {\em error} of 0.15~mag at least,  failing
the specified requirements. In the second case (right panel of
Figure~\ref{pancino_fig_red}), a small number (10) of red stars with deep
absorption bands are included in the SPSS set used to build a second dispersion
matrix. The second dispersion matrix is able to calibrate all red stars with
absorption features to  better than 1\%, exceeding the requirements.  

This example shows the importance of spectral features in the SPSS set used in
construction of the dispersion matrix. Hot stars have prominent absorption lines
in the blue, but no features in the red. The addition of a few red stars with
absorption bands ``trains'' the matrix in the  calibration of stars with features
in the red (effectively reducing degeneracy). Similarly, problems are encountered
in the calibration of emission line objects (peculiar hot stars and quasars, for
example). But it is quite difficult to include these objects into the SPSS set
since they are often variable.

Even if several types of objects are included when determining the dispersion
matrix, other effects can have a large impact on the degeneracy, such as edge
effects. For these reasons, the accurate choice of the SPSS set is crucial, but
does not solve the problem of degeneracy once and for all.  

\begin{figure}
\plottwo{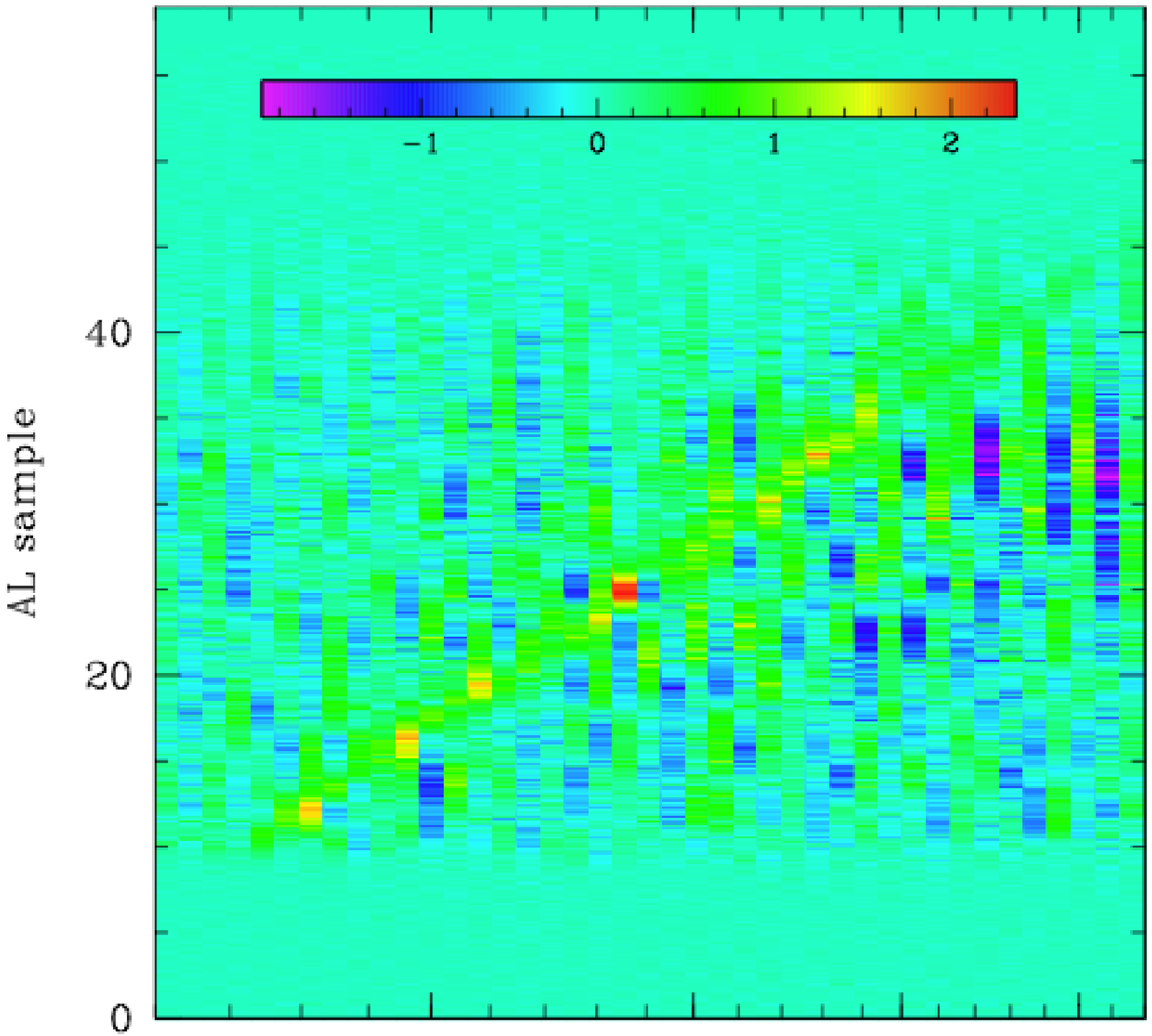}{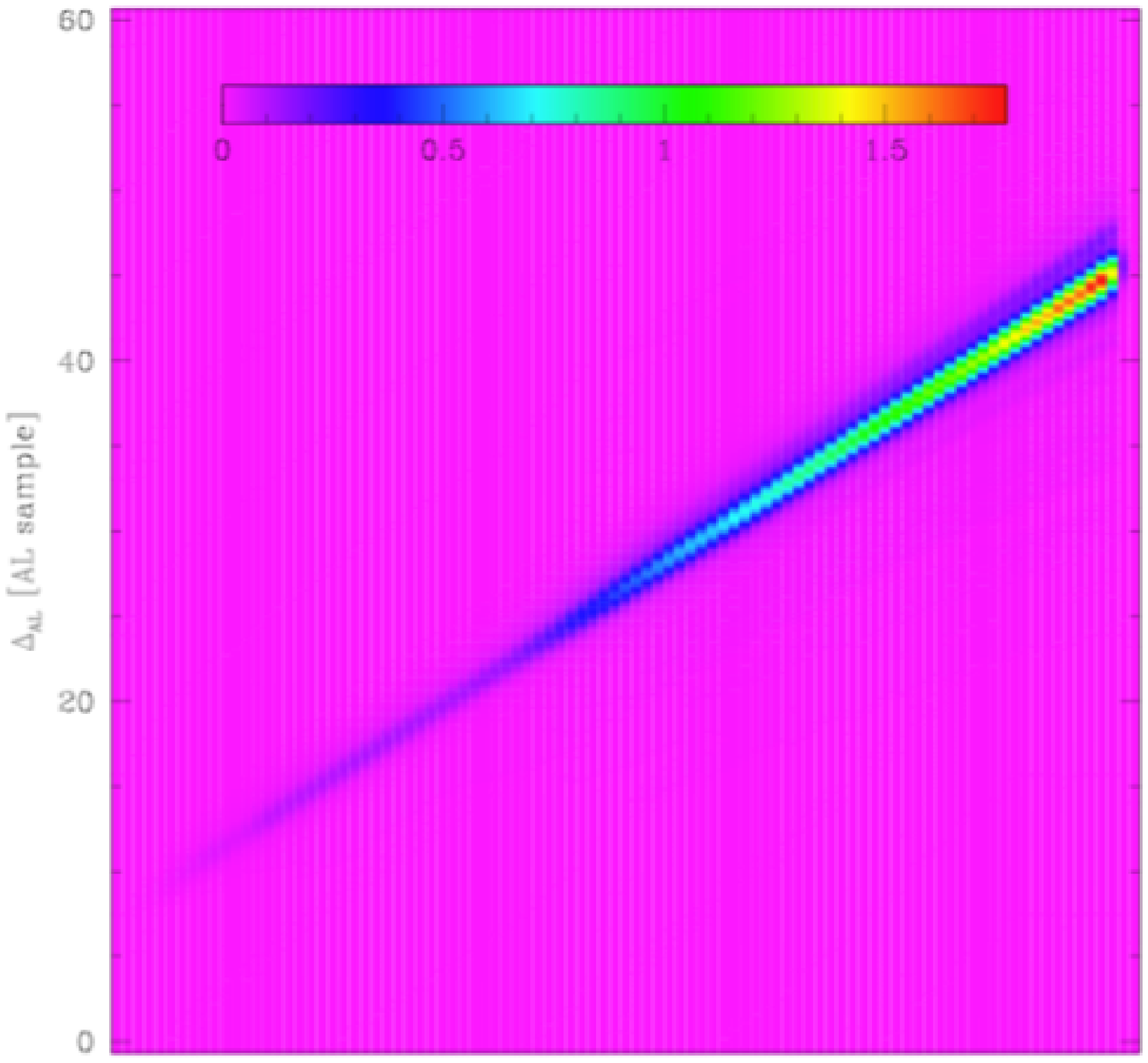}
\caption{Left: example of a degenerate dispersion matrix, D$_e$ (which is square,
see text), where the diagonal is drowned into noise-like patterns due to the
degenerate (non independent) set of SPSS used to construct it. Right: an example
of a diagonal dispersion matrix, obtained with an appropriate SPSS set and with
the use of a nominal dispersion matrix to further reduce degeneracy (see
text).\copyright ESA}
\label{pancino_fig_deg}
\end{figure}

\subsection{Nominal dispersion matrix}

To further reduce degeneracy of the effective dispersion matrix D$_e$, we can use
other constraints such as the fact that we know most aspects of the instrument
from pre-launch characterization. These include the quantum efficiency of the
CCDs, the optical layout and transmissivity, the nominal LSF at various positions
along the focal plane and at different wavelengths, the nominal dispersion
function and its variation along the focal plane. The slow change of these with
time can also be monitored to a certain extent, and included in the modelization.

We can therefore separate the dispersion matrix in a part that is theoretically
modeled based on pre-launch  instrument description and on its (partially
reconstructed) variation with time, which we call D$_n$ or {\em nominal dispersion
matrix}, and in a part that is completely unknown, which can be considered as a
correction matrix K, made of the residual corrections after the nominal model is
taken into account (Montegriffo et al. 2010)

$$ D_e = K \times D_n $$

The nominal matrix will be clean: diagonal and non-degenerate (see
Figure~\ref{pancino_fig_deg}). The  correction matrix will be partially
degenerate, but all signal that lies far away from the diagonal can be safely
considered spurious (the system varies in a continuous way, the corrections must
be ``small'' compared to the nominal system), and the part of the correction 
matrix close to its diagonal can be easily modeled.

To summarize all the previously  defined steps, once an appropriate SPSS set is
chosen, the calibration model becomes

$$ S_{obs} = D_e \times S_{smooth} = K \times D_n \times S_{smooth} $$

and the matrices involved can be easily inverted to calibrate all Gaia
observations since they are all square and (almost completely) diagonal.

\subsection{Integrated magnitudes}

A classical approach can be adopted for the absolute flux calibration of 
integrated M$_G$, M$_{BP}$, and M$_{RP}$ magnitudes (Ragaini et al., 2009a,b) in
the form 

$$ M = m + ZP $$

where M is the calibrated magnitude, m the internally calibrated one observed by
Gaia, and ZP is the required zero-point. No significant color term appears
necessary.  

However, if we consider that an integrated magnitude M is the convolution of the
spectral distribution S$_{true}$ and the effective passband B, we can calibrate
integrated magnidutes with the same approach adopted for BP/RP spectra, with a
much more homogeneous procedure from the point of view of pipeline code writing

$$ M = S_{true} \times B $$

Since generally the passband B is sampled differently than the SPSS flux table
S$_{true}$, we must apply smoothers to one or both S$_{true}$ and B. Similarly to
the case of BP/RP spectra, we can split B into two components

$$ B = K' \times B_n $$

where K' is a correction vector, made of the actual residuals to a thoretically
known -- or {\em nominal} -- effective passband B$_n$, known before launch and
slowly varying with time due to several causes, the most important one being the
CCDs quantum efficiency decreased due to radiation damage. With this kind of
treatment, the problem becomes a simple least quare fitting problem to derive the
unknown K' vector (Ragaini et al., 2010, in preparation). 

\subsection{RVS calibration}

The possibility of calibrating in flux the RVS spectra has been so far considered
a secondary problem, since both radial velocities and astrophysical parameters can
be derivad without the need of an absolute flux scale attached to the spectra. A
preliminary set of considerations (Trager, 2010) shows that in principle the SPSS
grid for the calibration of Gaia G-band and BP/RP data, that is presently under
construction, should be sufficiently sampled to ensure a flux calibration of RVS
spectra as well. We expect the topic to  be further explored by CU6 (Spectroscopic
processing) in the near future, but we will not consider RVS spectra flux
calibration in this paper. 

\section{The Gaia grid of spectrophotometric standard stars} 
\label{pancino_sec_spss}

From the above discussion, it is clear that the Gaia SPSS grid has to be chosen
with great care. The Gaia SPSS, or better their reference flux tables
(corresponding to S$_{true}$ in the previous Sections) should conform to the
following general requirements (van Leeuven et al. 2010):

\begin{itemize}
\item{Resolution R=$\lambda/\delta\lambda \simeq 1000$, i.e., they should 
oversample the Gaia BP/RP resolution by a factor of  4--5 at least;}
\item{Wavelength coverage: 330--1050~nm;}
\item{Typical uncertainty on the absolute flux scale, with respect to the assumed
calibration of Vega, of a few percent, excluding small troubled areas in the
spectral range (telluric bands residuals, extreme red and blue edges), where it
can be somewhat worse.}
\end{itemize}

The total number of SPSS in the Gaia grid should be of the order of 200--300
stars, including a variety of spectral types. Clearly, no such large and
homogeneous dataset exists in the literature yet\footnote{The CALSPEC database
(Bohlin, 2007) is not large enough for our purpose, especially considering the
strict criteria described below. Its extension to more than 100 SPSS is eagerly
awaited, but still not available to the public.}. It is therefore necessary to
build the Gaia SPSS grid with new, dedicated observations. We describe the
characteristics of the Gaia SPSS and of the dedicated observing campaigns in the
following Sections.

\begin{figure}
\plottwo{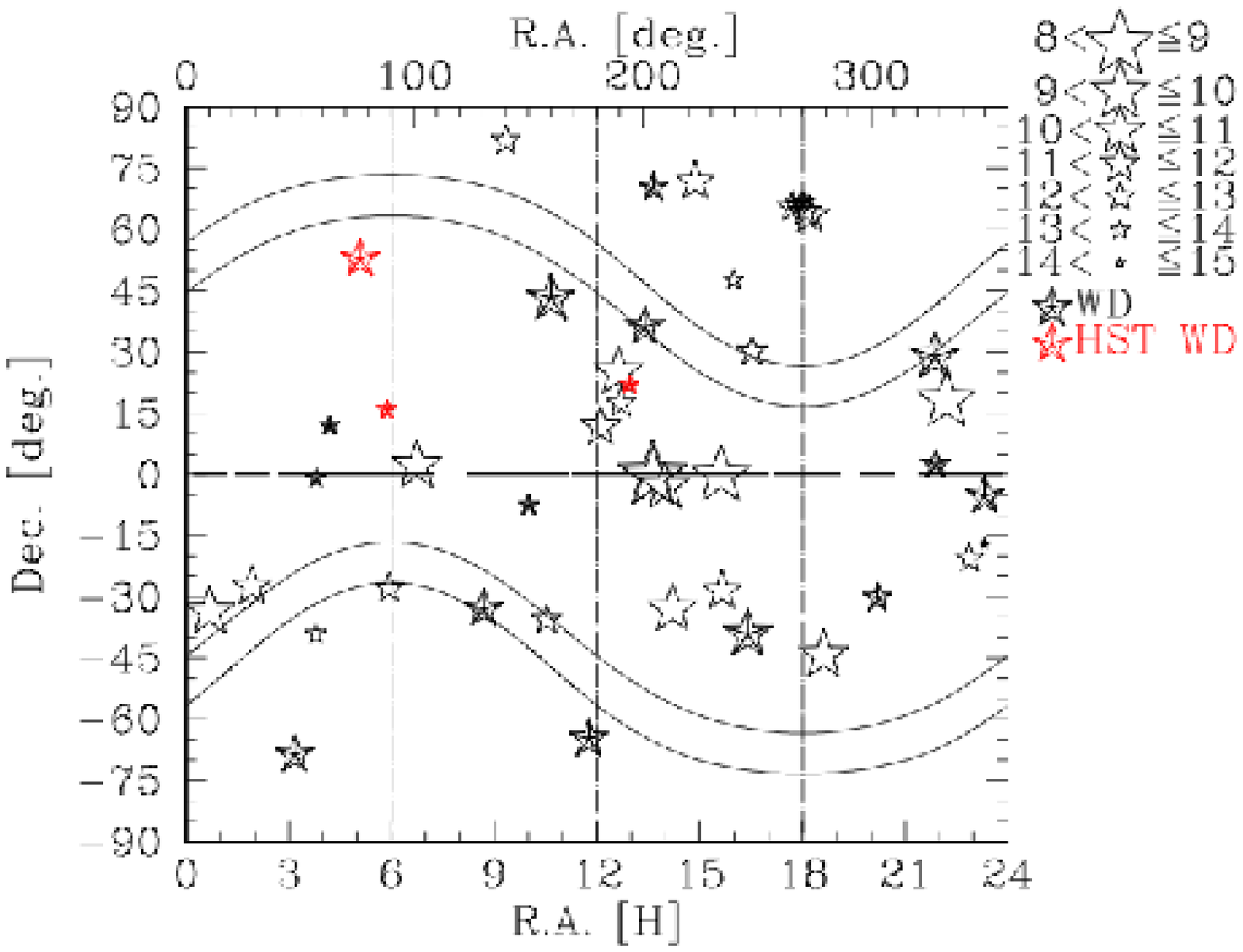}{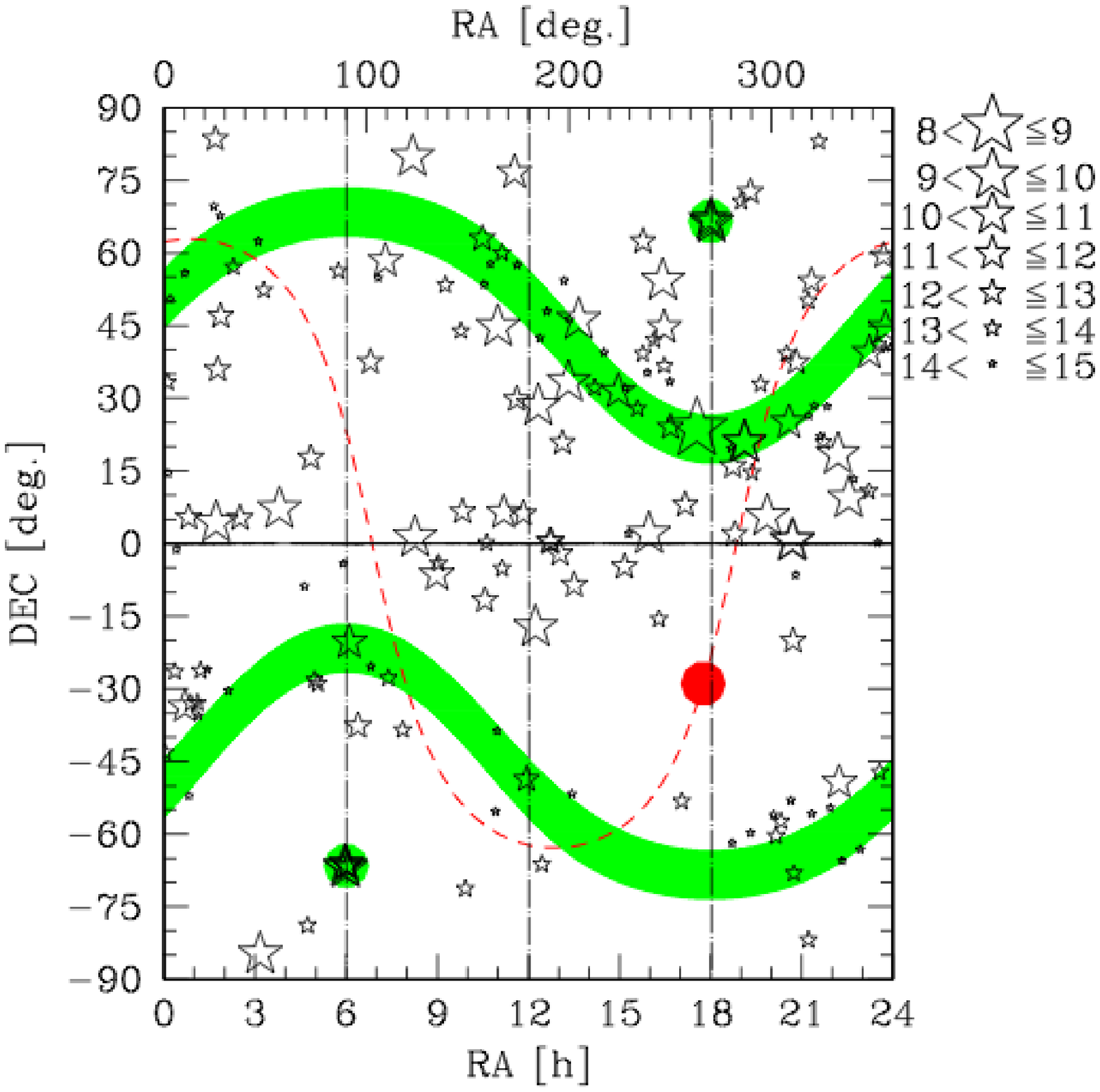}
\caption{The Ra/Dec distribution of Primary (left) and Secondary SPSS candidates
(right). The three Pillars are marked in red in the left panel, while the targets
close to the ecpliptic poles are marked in green in the right panel. The size of
symbols is inversely proportional to the SPSS magnitude. The two stripes at
$\pm$45~deg from the Ecliptic Poles are marked in both panels. \copyright ESA}
\label{pancino_fig_primary}
\end{figure}

\subsection{SPSS Candidates}

\begin{figure}
\epsscale{0.5}
\plotone{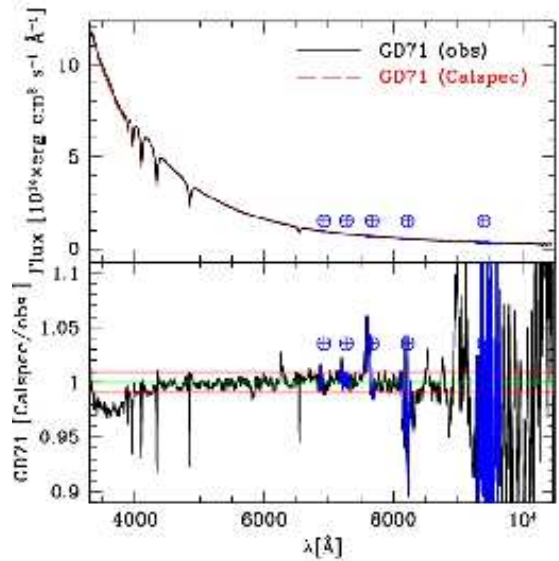}
\caption{The preliminary reductions (no telluric correction, only TNG
observations, library exctinction curve, and so on) of star GD~71 (top panel) are
compared with the CALSPEC spectrum (bottom panel). Prominent telluric features are
marked in both panels. Except for the spectral edges -- which will need to be
reconstructed with the use of models -- the main body of the spectrum is always
close to the CALSPEC one within 1\% or better. \copyright ESA}
\label{pancino_fig_gd71}
\end{figure}

We have followed a two steps approach (Bellazzini et al. 2007) that firstly
creates a set of {\em Primary SPSS}, i.e., well known SPSS that are calibrated
on the three {\em Pillars} of the
CALSPEC\footnote{http://www.stsci.edu/hst/observatory/cdbs/calspec.html} set,
described in Bohlin et al. (1995,2007), and tied to the Vega flux calibration by
Bohlin \& Gilliland (2004) and Bohlin et al. (2007). An example of the kind of
spectra obtained for Pillar GD~71 (with DoLoRes@TNG) is shown on
Figure~\ref{pancino_fig_gd71}. The Primary SPSS will constitute the ground-based
calibrators of the actual Gaia grid, and need to conform to the following
requirements (van Leeuwen 2010):

\begin{itemize}
\item{Primary SPSS have spectra as featureless as possible;}
\item{Primary SPSS shall be validated against variability;}
\item{Primary SPSS have already well known SEDs;}
\item{The magnitude of each Primary SPSS grants a resulting S/N$\simeq$100
per pixel over most of the wavelength range when observed from the ground with 2m
class telescopes;}
\item{The location of Primary SPSS is in non crowded areas of the sky;}
\item{Primary SPSS cover a range of RA and Dec to ensure all-year-round ground based
observations from both hemispheres.}
\end{itemize}

The Primary SPSS candidates set is described in more detail in Altavilla et al.
(2008), and some of the most important sources for Primary candidates are the
CALSPEC grid, Oke (1990), Hamuy et al. (1992,1994), Stritzinger et al. (2005)
and others. The actual Gaia SPSS grid, or {\em Secondary SPSS}, conforms to a
different set of requirements (van Leeuwen 2010):

\begin{itemize}
\item{Secondary SPSS have spectra as featureless as possible (but see below
for exceptions);}
\item{Secondary SPSS shall be validated against variability;}
\item{The magnitude and sky location (i.e., number of useful, clean transits,
see Carrasco et al. 2006, 2007) of Secondary SPSS grants a resulting
S/N$\simeq$100 per sample over most of the wavelength range when observed by
Gaia (end of mission);}
\item{Secondary SPSS cover a range of spectral types and spectral shapes, as
needed to ensure the best possible claibration of all kinds of objects observed by
Gaia.}
\end{itemize}

As already mentioned, Secondary SPSS will be mostly hot and featureless stars
but will include a small number of selected spectral types, to ensure that the
calibration model can work on all objects types. More details on Secondary SPSS
can be found in Altavilla et al. (2010), including a long list of literature
calatalogues and online databases from which the candidates are extracted.
Clearly, all the Primary SPSS that, at the end of the data reductions, will
satisfy also the criteria for Secondary SPSS, will be included in the Gaia SPSS
grid.

Additional, special members of the Secondary SPSS candidates are: (1) a few
selected SPSS around the Ecliptic Poles, two regions of the sky  that will be
repeatedly observed by Gaia, in the first two weeks after reaching its orbit in
L2, for calibration purposes; (2) a few M stars with deep absorption features in
the red; (3) a few SDSS stars that have been observed in SEGUE sample (Yanny et
al., 2009), since the SEGUE sample has the potential of being extremely useful
in the Gaia flux calibration (Bellazzini et al. 2010), both internal (relative)
and external (absolute); (4) a few well known SPSS that are among the targets of
the ACCESS mission (Kayser et al., 2010), dedicated to the absolute flux
measurement of a few stars besides Vega.

\begin{figure}
\plottwo{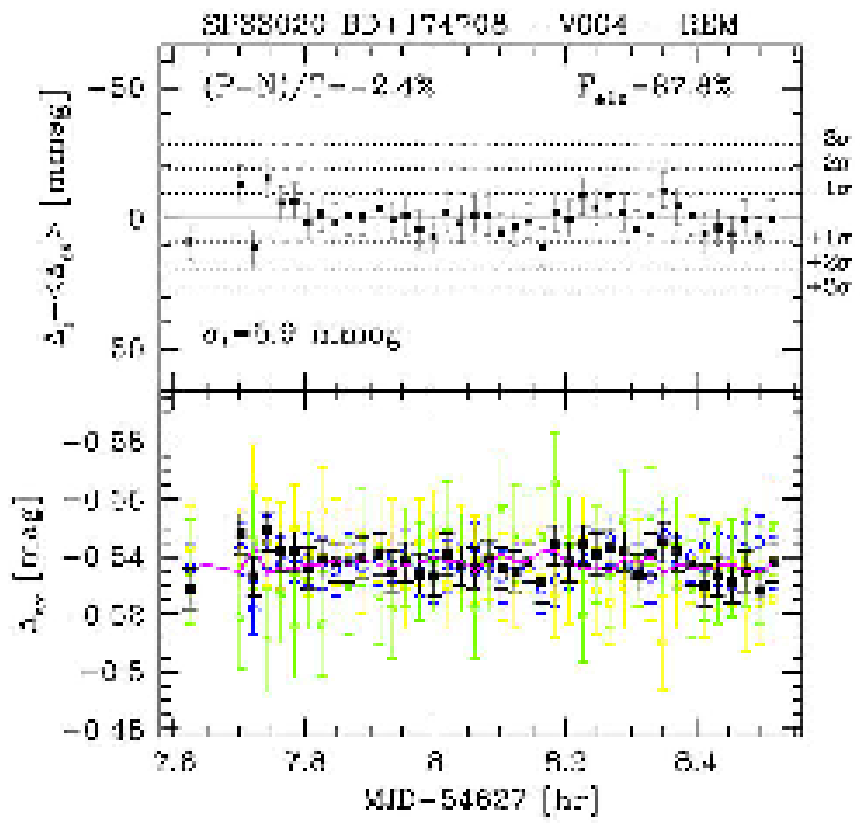}{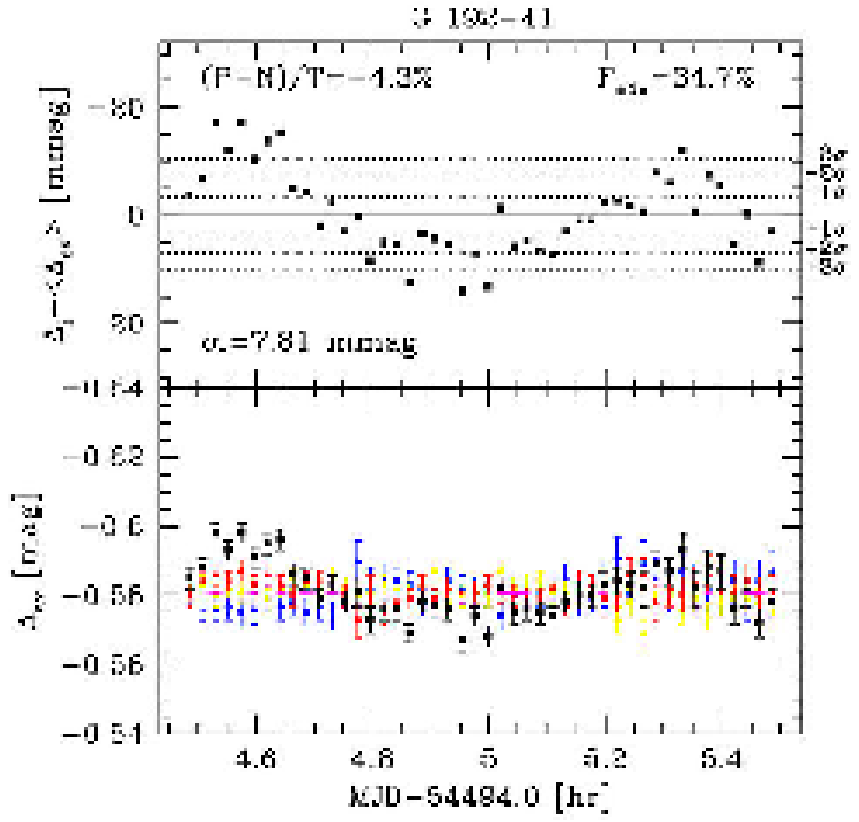}
\caption{Left: example of a short-term variability curve for a constant SPSS
candidate. Right: example of a short-term variability curve for a variable SPSS
candidate. \copyright ESA}
\label{pancino_fig_var}
\end{figure}

\subsection{Observation strategy and campaigns}

A basic consideration when starting the observations of such a large campaign, is
that the traditional spectrophotometry techniques require too much observing time:
each SPSS should be spectroscopically observed in perfectly photometric
conditions, ideally more than once. No TAC (Time Allocation Committee) would grant
such a large amount of observing time to a proposal that does not contain any
cutting edge science in it.  

We therefore chose (Bellazzini et al. 2007) to split the  problem into two parts:
(1) spctra are taken in good  sky conditions, but not necessarily perfectly
photometric; they are calibrated with the help of a {\em Pillar} or {\em Primary
SPSS} thus recovering the correct spectral {\em shape}; (2) absolute photometry in
the B, V, and R (sometimes I) Johnson-Cousins bands is taken in photometric sky
conditions and used to fix the spectral {\em zero-point} by means of synthetic
photometry. This is motivated by the fact that absolute photometric night points
are faster to obtain than spectra. A subset of SPSS candidates will be
spectroscopically observed in photometric sky conditions, to check the whole
procedure. 

Besides the {\em Main Campaign} just described, it is necessary to monitor
candidate SPSS for constancy ({\em Auxiliary Campaign}), since very few of them
have systematically been monitored in  the literature, and there are illustrious
examples of stars that showed unexpected variability (Landolt \& Uomoto, 2007).
An example of a different kind of problem, that could greatly benefit from good
quality dedicated imaging, is star HZ~43. It was initially chosen by Bohlin et
al. (1995) to be one of the Pillars, and later rejected because of an optical
companion lying 3" away, only visible in the V band, that made it useless as an
SPSS from the ground. 

Most of our SPSS candidates are WD close to the instability strip, and sometimes
have poorly known magnitudes, so it is necessary to monitor them for short-term
variability on 1--2~h (Figure~\ref{pancino_fig_var}). Binary systems are
frequent and can be found at all spectral types, so we also monitor all our
candidates for long-term variability (3~yrs) collecting approximately 4 night
points per year. These two monitoring campaigns rely on relative photometry
(using stars in the field of view) to derive variability curves. An SPSS
candidate is considered constant if it does not vary with an amplitude larger
than a few mmag.

The facilities that are being used for the two observing campaigns are (Federici
et al., 2007, Altavilla et al. 2010):

\begin{itemize}
\item{EFOSC@NTT, La Silla, Chile (primarily {\em Main campaign});}
\item{ROSS@REM, La Silla, Chile (primarily {\em Auxiliary campaign});}
\item{LARUCA@1.5m, San Pedro Martir, Mexico (primarily {\em Auxiliary campaign}
and absolute photometry);}
\item{BFOSC@Cassini, Loiano, Italy (primarily {\em Auxiliary campaign});}
\item{CAFOS@2.2m, Calar Alto, Spain (primarily {\em Main campaign});}
\item{DoLoRes@TNG, La Palma, Spain (primarily {\em Main campaign});}
\end{itemize}

Observations started in the second half of 2006, comprising more than 35
accepted proposals. We have been awarded a total of 230 observing nights
approximately, at the rate of 33 per semester. More than 50\% of this time
resulted in at least partially useful data. Given the large number of facilities
involved, and of different observers, it has become necessary to establish
strict observing protocols (Pancino et al. 2008,2009). The campaings are now
more than 50\% complete, with the spectroscopy observations 75\% complete, and
we expect to complete all our observing campaigns around 2013. 

\begin{figure}
\plotone{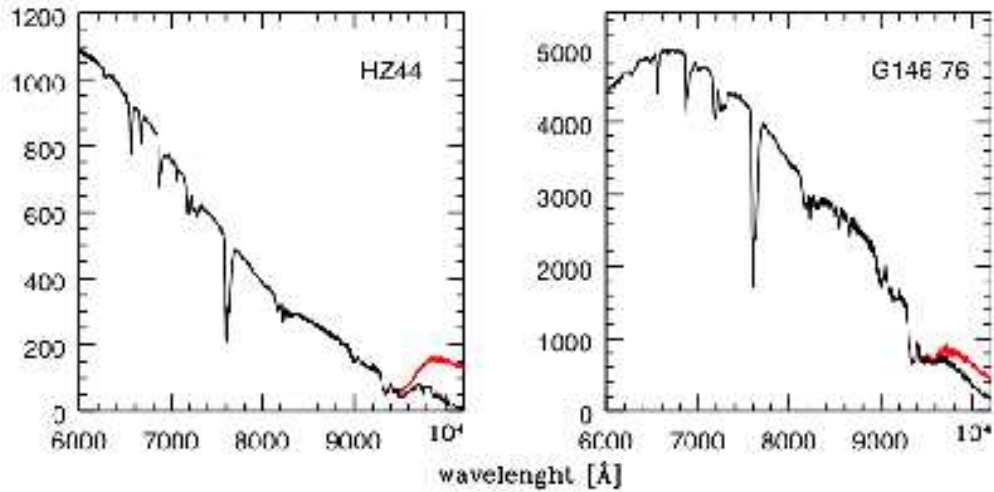}
\caption{Example of the 2$^{nd}$ order contamination in the DoLoRes grism LR-R,
and of its correction for stars HZ~44 and G~146--76. The solid black curves are
the corrected spectra while the red curves are the contaminating light coming from
the 2$^{nd}$ order dispersed blue light. \copyright ESA}
\label{pancino_fig_2nd}
\end{figure}

\subsection{Data reduction and analysis}

The large amount of data collected needs to be reduced and analysed with the
maximum possible precision and homogeneity. An initial set of data is collected
for each CCD/instrument/telescope combination and an  ``Instrument
Familiarization Plan'' (IFP) is conducted, to derive shutter times, linearity,
calibration frames and lamps stability, photometric distortions, 2$^{nd}$ order
contamination of spectra (Figure~\ref{pancino_fig_2nd}), and so on. This plan is
now almost complete, and the protocols are presently being finalized and
written.

The data reduction is regulated by strict data reduction protocols, that are
presently being finalized. While the data reduction methods are fairly standard,
care must be taken in considering the characteristics of each instrument as
determined during the IFP, to extract the highest possible quality from each
instrument. Semi-automatic quality check (QC) criteria are defined for each kind
of observation (minimum/maximum S/N, seeing and roundness requiremets of images,
presence of bad columns, companions, and so on). Only frames that pass the QC
are reduced. For imaging, we term ``data reduction'' the removal of the
instrument characteristics (dark, bias, flat-field, fringing), QC, and the
measurement of aperture photometry with SExtractor (Bertin \& Arnouts, 1996).
The data products are 2D reduced images and aperture magnitude catalogues. For
spectroscopy, we term ``data reduction'' the removal of instrumental features
(dark, bias, flat-field, illumination correction, wavelength calibration,
2$^{nd}$ order contamination correction, relative flux calibration, telluric
features removal), followed by QC and spectra extraction. The data products are
2D reduced frames, 1D extracted and wavelength calibrated spectra, 1D flux
calibrated spectra, 1D telluric absorption corrected spectra.

The data reduction procedures are well advanced for photometry (almost half of the
data reduced) and are just started for spectra (10\% of the data reduced) at the
moment of writing. 

The data analysis is presently in the design and testing phases. The study of
short-term variability curves is proceeding (10\% of the data analysed, see
Figure~\ref{pancino_fig_var}). Absolute photometry and relative spectroscopy
procedures are being refined: for example, preliminary end-to-end reduction of
photometric imaging nights have been performed for TNG and NTT observations, to
allow us to identify those nights that were actually photometric and did not
need to be repeated. Preliminary extinction curves have been determined for TNG
and CAHA spectroscopic observations, allowing us to see that extinction varies
in a grey manner (within a few percent) even in the case of some Calima (desert
dust) in the sky in La Palma. 

The final data products for the {\em Auxiliary campaign} will be relative
magnitudes and lightcurves for all the monitored candidate SPSS; for the {\em
Main campaign}, absolute magnitudes and errors will be released together with
their uncertainties and flux tables (Figure~\ref{pancino_fig_gd71}) in the form
($\lambda$(nm),F$_{\lambda}$(photons s$^{-1}$ m$^{-2}$ nm$^{-1}$)). Possibly,
also other intermediate data products will be released (see above).

\begin{figure}
\plotone{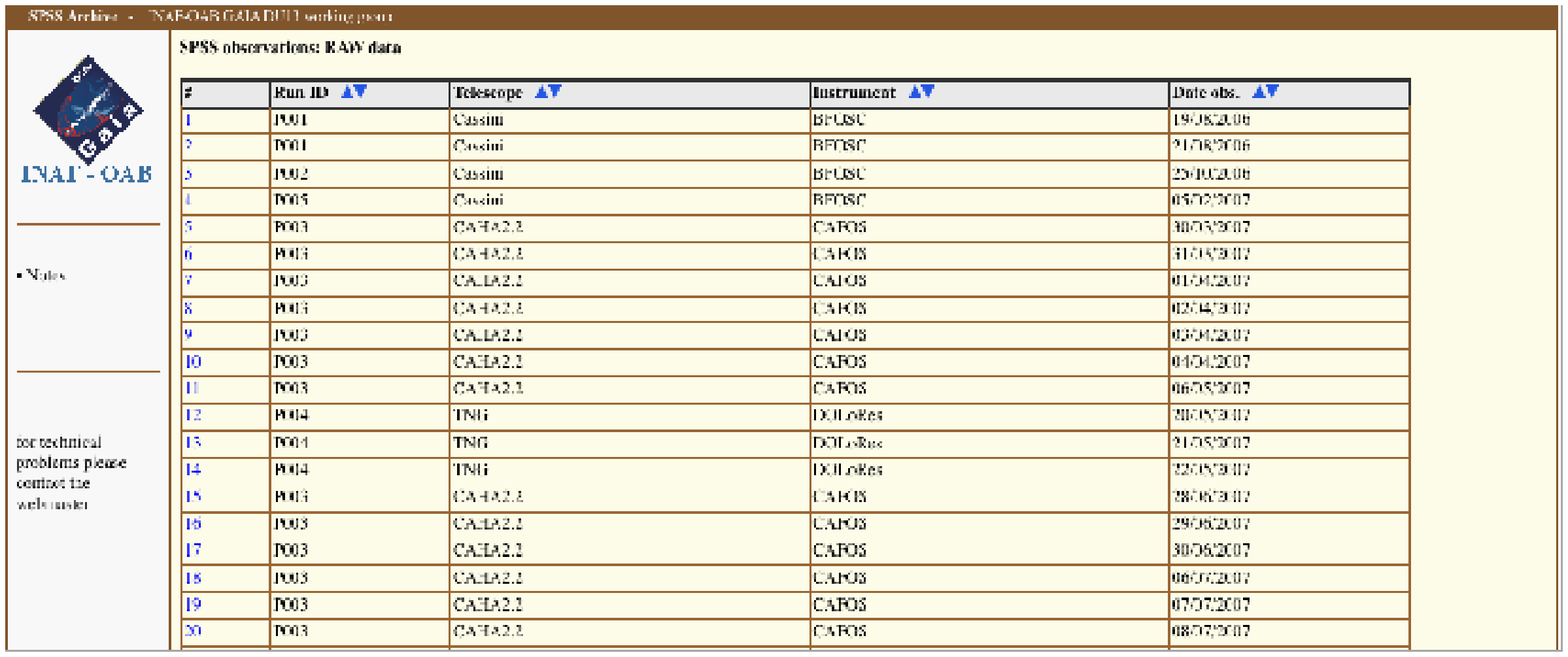}
\caption{The simple browsing interface of the Wiki-Bo local SPSS archive. This
snapshot refers to the {\em raw data archive}, a similar web page exists for
data products. \copyright ESA}
\label{pancino_fig_arch}
\end{figure}

\subsection{Data availability}

All the data coming from ground based observations of SPSS, along with the
collected literature information and measurements, are stored in the CU5-DU13
local Wiki pages in Bologna (Wiki-Bo)\footnote{http://yoda.bo.astro.it/wiki,
guest username and password can be obtained from E.~Pancino.}. Wiki-Bo contains
also all our technical documentation, internal reports, observation status and
data products, along with literature references and sources, observing proposals
and all the like. The raw and reduced data products are stored in a local
archive\footnote{http://spss.bo.astro.it, guest username and password can be
obtained from E.~Pancino.} for internal purposes
(Figure~\ref{pancino_fig_arch}). 

In the future, when CU9 will be started (Catalogue production and access) it is
foreseen that all the ground-based data that are used for the calibration of
Gaia data (radial velocity standards, SPSS, spectral libraries, Ecliptic pole
observations, observations of Gaia itself from the ground, and so on) will be
published as well, although no decision on the format and type of data products
has been taken yet. 

\section{Conclusions}

The Gaia mission and its data reduction is a challenging enterprise, carried on
by ESA and the European scientific community. As an example of the DPAC (Data
Processing and Analysis Consortium) tasks, I have briefly summarized the problem
of the external (absolute) flux calibration of (spectro)photometric Gaia data,
and more specifically of the BP/RP low resolution spectra and the integrated
G-band and BP/RP magnitudes. An innovative calibration model is presently under
study and testing, and a large ($\simeq200-300$) grid of SPSS with 1--3\% flux
calibration with respect to Vega is being built from multi-site ground-based
obseervations. 

But once the Gaia data will become available, a greater challenge will have to
be faced: the impact in almost all fields of astrophysics require that the
scientific community (and not only the European one) be adequately prepared to
extract the most scientific output from the data. The training of a new
generation of scientists, and the collection of complementary data, necessary to
answer key questions when combined with Gaia data, should start now. The
challenge requires that large groups of scientists get efficiently organized and
ready to collaborate on large and comprehensive datasets.

\end{document}